\documentclass[12pt]{article}
\usepackage{epsfig}
\usepackage{lineno}
\usepackage{amsmath}
\usepackage{hhline}
\usepackage{amssymb}
\usepackage{times}
\usepackage{cite}

\newlength{\dinwidth}
\newlength{\dinmargin}
\begin{document}
\newcommand{\ra}{\rightarrow}
\newcommand{\ccb}{c\bar{c}}
\newcommand{\bbb}{b\bar{b}}
\newcommand{\ptrel}{$\,p_t^{rel}\,$}
\newcommand{\ptrelm}{$\,p_t^{rel}$}
\newcommand{\del}{$\delta\;$}
\newcommand{\ptmu}{$p_t^{\mu}$}
\newcommand{\etamu}{$\eta^{\mu}$}
\newcommand{\xgobs}{$x_{\gamma}^{obs}\,$}
\newcommand{\xgobsm}{x_{\gamma}^{obs}}
\newcommand{\pom}{{I\!\!P}}
\newcommand{\reg}{{I\!\!R}}
\newcommand{\slowpi}{\pi_{\mathit{slow}}}
\newcommand{\fiidiii}{F_2^{D(3)}}
\newcommand{\fiidiiiarg}{\fiidiii\,(\beta,\,Q^2,\,x)}
\newcommand{\n}{1.19\pm 0.06 (stat.) \pm0.07 (syst.)}
\newcommand{\nz}{1.30\pm 0.08 (stat.)^{+0.08}_{-0.14} (syst.)}
\newcommand{\fiidiiiful}{F_2^{D(4)}\,(\beta,\,Q^2,\,x,\,t)}
\newcommand{\fiipom}{\tilde F_2^D}
\newcommand{\ALPHA}{1.10\pm0.03 (stat.) \pm0.04 (syst.)}
\newcommand{\ALPHAZ}{1.15\pm0.04 (stat.)^{+0.04}_{-0.07} (syst.)}
\newcommand{\fiipomarg}{\fiipom\,(\beta,\,Q^2)}
\newcommand{\pomflux}{f_{\pom / p}}
\newcommand{\nxpom}{1.19\pm 0.06 (stat.) \pm0.07 (syst.)}
\newcommand {\gapprox}
   {\raisebox{-0.7ex}{$\stackrel {\textstyle>}{\sim}$}}
\newcommand {\lapprox}
   {\raisebox{-0.7ex}{$\stackrel {\textstyle<}{\sim}$}}
\def\gsim{\,\lower.25ex\hbox{$\scriptstyle\sim$}\kern-1.30ex%
\raise 0.55ex\hbox{$\scriptstyle >$}\,}
\def\lsim{\,\lower.25ex\hbox{$\scriptstyle\sim$}\kern-1.30ex%
\raise 0.55ex\hbox{$\scriptstyle <$}\,}
\newcommand{\pomfluxarg}{f_{\pom / p}\,(x_\pom)}
\newcommand{\dsf}{\mbox{$F_2^{D(3)}$}}
\newcommand{\dsfva}{\mbox{$F_2^{D(3)}(\beta,Q^2,x_{I\!\!P})$}}
\newcommand{\dsfvb}{\mbox{$F_2^{D(3)}(\beta,Q^2,x)$}}
\newcommand{\dsfpom}{$F_2^{I\!\!P}$}
\newcommand{\gap}{\stackrel{>}{\sim}}
\newcommand{\lap}{\stackrel{<}{\sim}}
\newcommand{\fem}{$F_2^{em}$}
\newcommand{\tsnmp}{$\tilde{\sigma}_{NC}(e^{\mp})$}
\newcommand{\tsnm}{$\tilde{\sigma}_{NC}(e^-)$}
\newcommand{\tsnp}{$\tilde{\sigma}_{NC}(e^+)$}
\newcommand{\st}{$\star$}
\newcommand{\sst}{$\star \star$}
\newcommand{\ssst}{$\star \star \star$}
\newcommand{\sssst}{$\star \star \star \star$}
\newcommand{\tw}{\theta_W}
\newcommand{\sw}{\sin{\theta_W}}
\newcommand{\cw}{\cos{\theta_W}}
\newcommand{\sww}{\sin^2{\theta_W}}
\newcommand{\cww}{\cos^2{\theta_W}}
\newcommand{\trm}{m_{\perp}}
\newcommand{\trp}{p_{\perp}}
\newcommand{\trmm}{m_{\perp}^2}
\newcommand{\trpp}{p_{\perp}^2}
\newcommand{\alp}{\alpha_s}
\newcommand{\alps}{\alpha_s}
\newcommand{\sqrts}{$\sqrt{s}$}
\newcommand{\LO}{$O(\alpha_s^0)$}
\newcommand{\Oa}{$O(\alpha_s)$}
\newcommand{\Oaa}{$O(\alpha_s^2)$}
\newcommand{\PT}{p_{\perp}}
\newcommand{\JPSI}{J/\psi}
\newcommand{\sh}{\hat{s}}
\newcommand{\uh}{\hat{u}}
\newcommand{\MP}{m_{J/\psi}}
\newcommand{\PO}{I\!\!P}
\newcommand{\xbj}{x}
\newcommand{\xpom}{x_{\PO}}
\newcommand{\ttbs}{\char'134}
\newcommand{\xpomlo}{3\times10^{-4}}
\newcommand{\xpomup}{0.05}
\newcommand{\dgr}{^\circ}
\newcommand{\pbarnt}{\,\mbox{{\rm pb$^{-1}$}}}
\newcommand{\gev}{\,\mbox{GeV}}
\newcommand{\WBoson}{\mbox{$W$}}
\newcommand{\fbarn}{\,\mbox{{\rm fb}}}
\newcommand{\fbarnt}{\,\mbox{{\rm fb$^{-1}$}}}
\newcommand{\qsq}{\ensuremath{Q^2} }
\newcommand{\gevsq}{\ensuremath{\mathrm{GeV}^2} }
\newcommand{\et}{\ensuremath{E_t^*} }
\newcommand{\rap}{\ensuremath{\eta^*} }
\newcommand{\gp}{\ensuremath{\gamma^*}p }
\newcommand{\dsiget}{\ensuremath{{\rm d}\sigma_{ep}/{\rm d}E_t^*} }
\newcommand{\dsigrap}{\ensuremath{{\rm d}\sigma_{ep}/{\rm d}\eta^*} }

\newcommand{\bul}{$\,$ -- }

\newcommand{\icaption}[1]{\caption{\it #1}}

\def\Journal#1#2#3#4{{#1} {\bf #2} (#3) #4}
\def\NCA{\em Nuovo Cimento}
\def\NIM{\em Nucl. Instrum. Methods}
\def\NIMA{{\em Nucl. Instrum. Methods} {\bf A}}
\def\NPB{{\em Nucl. Phys.}   {\bf B}}
\def\PLB{{\em Phys. Lett.}   {\bf B}}
\def\PRL{\em Phys. Rev. Lett.}
\def\PRD{{\em Phys. Rev.}    {\bf D}}
\def\ZPC{{\em Z. Phys.}      {\bf C}}
\def\EJC{{\em Eur. Phys. J.} {\bf C}}
\def\CPC{\em Comp. Phys. Commun.}

\begin{titlepage}

\begin{flushleft}
DESY 05-004 \hfill ISSN 0418-9833 \\
January 2005
\end{flushleft}

\vspace{2cm}

\begin{center}
\begin{Large}

{\bf Measurement of Beauty Production
    at HERA \\ Using Events with Muons and Jets}

\vspace{2cm}

H1 Collaboration

\end{Large}
\end{center}

\vspace{2cm}
%
%
%
%

\begin{abstract}
\noindent
A measurement of the beauty production cross section
in $ep$ collisions at a centre-of-mass energy of 319 GeV 
is presented.
The data were collected with the H1 detector at the HERA collider 
in the years 1999-2000. 
%
Events are selected by requiring the presence of jets and muons
in the final state.
Both the long lifetime and the large mass of $b$-flavoured hadrons 
are exploited to identify events containing beauty quarks.
Differential cross sections are measured in photoproduction,
with photon virtualities \mbox{$Q^2<1$ GeV$^2$},
and in deep inelastic scattering, where \mbox{$2<Q^2<100$ GeV$^2$}.
The results are compared with perturbative QCD calculations 
to leading and next-to-leading order.
The predictions are found to be somewhat lower than the data.
\end{abstract}

\vspace{1.5cm}

\begin{center}
To be submitted to {\it Eur.\ Phys.\ J.\ C}
\end{center}

\end{titlepage}
\begin{flushleft}

A.~Aktas$^{10}$,               
V.~Andreev$^{26}$,             
T.~Anthonis$^{4}$,             
S.~Aplin$^{10}$,               
A.~Asmone$^{34}$,              
A.~Babaev$^{25}$,              
S.~Backovic$^{31}$,            
J.~B\"ahr$^{39}$,              
A.~Baghdasaryan$^{38}$,        
P.~Baranov$^{26}$
% ...

\bigskip{\it
 $ ^{1}$ I. Physikalisches Institut der RWTH, Aachen, Germany$^{ a}$ \\
 $ ^{2}$ III. Physikalisches Institut der RWTH, Aachen, Germany$^{ a}$ \\
% ...
$ ^{4}$ Inter-University Institute for High Energies ULB-VUB, Brussels;
          Universiteit Antwerpen, Antwerpen; Belgium$^{ c}$ \\
% ...
$ ^{10}$ DESY, Hamburg, Germany \\
% ...
$ ^{25}$ Institute for Theoretical and Experimental Physics,
          Moscow, Russia$^{ l}$ \\
 $ ^{26}$ Lebedev Physical Institute, Moscow, Russia$^{ e}$ \\
% ...
$ ^{31}$ Faculty of Science, University of Montenegro,
          Podgorica, Serbia and Montenegro \\
% ...
$ ^{34}$ Dipartimento di Fisica Universit\`a di Roma Tre
          and INFN Roma~3, Roma, Italy \\
% ...
$ ^{38}$ Yerevan Physics Institute, Yerevan, Armenia \\
 $ ^{39}$ DESY, Zeuthen, Germany \\
% ...
$ ^a$ Supported by the Bundesministerium f\"ur Bildung und Forschung, FRG,
      under contract numbers 05 H1 1GUA /1, 05 H1 1PAA /1, 05 H1 1PAB /9,
      05 H1 1PEA /6, 05 H1 1VHA /7 and 05 H1 1VHB /5 \\
% ...
$ ^c$ Supported by FNRS-FWO-Vlaanderen, IISN-IIKW and IWT
      and  by Interuniversity
Attraction Poles Programme,
      Belgian Science Policy \\
% ...
$ ^e$ Supported by the Deutsche Forschungsgemeinschaft \\
% ...
$ ^l$ Partially Supported by Russian Foundation
      for Basic Research, grant    no. 00-15-96584 \\
}

\end{flushleft}

\newpage
\pagestyle{plain}
\section{Introduction}
A measurement is presented of open beauty production 
$ep \rightarrow eb\bar{b}X$ in $ep$ collisions with the H1 
detector at HERA.
The measurement spans the kinematic range from the domain of
photoproduction, in which the exchanged photon is quasi-real
($Q^2 \sim 0$), to the region of electroproduction,
or deep inelastic scattering (DIS), with photon virtualities \mbox{$2<Q^2<100$~GeV$^2$}.
For beauty production, calculations in perturbative 
quantum chromodynamics (pQCD) are expected to give reliable predictions, 
as the mass $m_b$ of the $b$ quark \mbox{($\,m_b \sim$ 5 GeV)}
provides a hard scale.
With the phase space covered in this analysis the interplay of 
the hard scales $m_b$, $Q^2$ and the transverse momenta of the $b$ quarks
can be probed.

%
First measurements of the beauty cross section at 
HERA~\cite{Adloff:1999nr,Breitweg:2000nz} 
were higher than pQCD predictions calculated at next-to-leading order (NLO).
Similar observations were made in hadron-hadron collisions~\cite{hadronb}
and also in two-photon interactions~\cite{Acciarri:2000kd}.
Recent beauty production
measurements from the H1~\cite{Aktas:2004az} and
ZEUS Collaborations~\cite{Chekanov:2004xy} are in 
better agreement with QCD predictions or again somewhat 
higher~\cite{Chekanov:2004tk}.

In this paper, photoproduction events with at least two jets ($jj$)
and a muon ($\mu$) in the final state are used
to measure the beauty cross section for
$
ep \rightarrow e b\bar{b} X \rightarrow e jj \mu X'
$.
In deep inelastic scattering, the process
$
ep \rightarrow e b\bar{b} X \rightarrow e j \mu X'
$
is measured with at least one jet and a muon in the final state.
For the first time at HERA, two distinct features of $B$-hadrons
are exploited simultaneously to discriminate events containing beauty from
those with only charm or light quarks: the large mass
and the long lifetime.
The $B$-hadron mass leads to a broad distribution of the 
transverse momentum \ptrel of 
decay muons relative to the beauty quark jet direction.
The $B$-hadron lifetime is reflected in the large 
impact parameters \mbox{$\delta \sim 200~\mu$m} of 
decay muon tracks relative to the primary vertex. 
The precision measurement
of muon track impact parameters is made possible by
the H1 Central Silicon Track detector~\cite{Pitzl:2000wz}.
The fractions of $b$ quark events in the data samples 
are determined from a fit to the two-dimensional distribution of 
\ptrel and\,\,$\delta$.

This paper is organized as follows.
In section \ref{sec:qcd}, an introduction to the physics of
beauty production in $ep$ collisions is given.
The relevant features of the
H1 detector are described 
in section \ref{sec:det}. Section \ref{sec:sel} describes
the event selection. The Monte Carlo simulations
and NLO QCD calculations are presented in sections 
\ref{sec:mc} and \ref{sec:nlo}.
Comparisons of the data samples with the Monte Carlo simulations
are shown in section \ref{sec:control}.
The fit procedure used to determine the 
$b$-fraction and the systematic errors of the measurement
are explained in sections \ref{sec:fit} and \ref{sec:syserr}.
Finally, the results are presented in section \ref{sec:results} and
conclusions are drawn in section~\ref{sec:conclusions}.

\section{Heavy Quark Production in {\boldmath $ep$} Collisions}
\label{sec:qcd}
In pQCD, at leading order, two distinct classes of processes contribute to 
the production of beauty quarks in $ep$ collisions at HERA.
\begin{figure}
\setlength{\unitlength}{1cm} 
\begin{picture}(14.0,5.0)
\put(1.,0.5){\epsfig{file=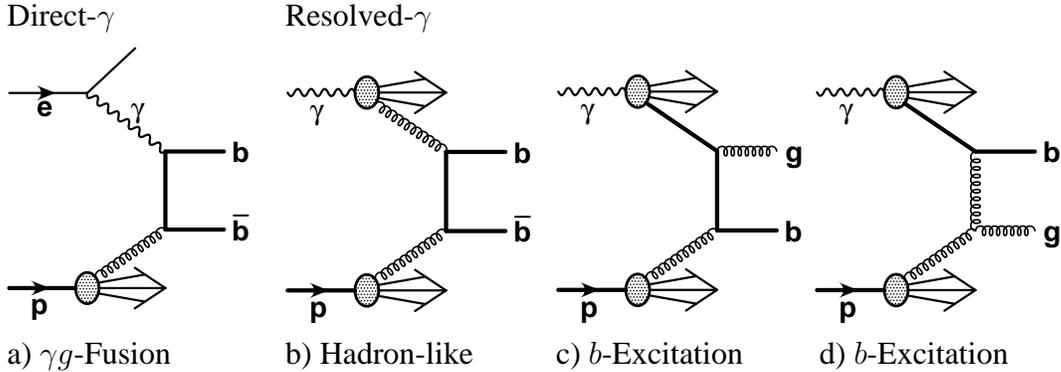,width=14cm}}
\put(1.,4.5){Direct-$\gamma$}
\put(4.7,4.5){Resolved-$\gamma$}
\put(1.,0.){a) $\gamma g$-Fusion}
\put(4.7,0.){b) Hadron-like}
\put(8.3,0.){c) $b$-Excitation}
\put(11.8,0.){d) $b$-Excitation}
\end{picture}
\caption{Beauty production processes in leading order pQCD.}
\label{fig:feynman}
\end{figure} 
In direct-photon processes (figure~\ref{fig:feynman}a),
the photon emitted from the positron enters
the hard process $\gamma g \rightarrow b\bar{b}$ directly.
In resolved-photon processes (figures~\ref{fig:feynman}b to d), 
the photon fluctuates into a hadronic state before the hard 
interaction and acts as a source of partons, one of 
which takes part in the hard interaction. 
Resolved photon processes are expected to
contribute significantly in the photoproduction
region, in which the photon is quasi-real, 
and to be suppressed towards higher $Q^2$.

For the kinematic range covered in this analysis,
the majority of events are in the region 
$p_{t,b} \gtrsim m_b$, where $p_{t,b}$ is the momentum of the
outgoing $b$ quark transverse to the photon-proton axis in the 
photon-proton centre-of-mass frame.
In this region, NLO calculations in the massive scheme~\cite{Frixione:1995qc,Harris:1995dv}
are expected to give reliable results. In this 
scheme,  $u$, $d$ and $s$ are the only active flavours in the proton 
and the photon, and charm and beauty are produced dynamically in the hard scattering.
At large transverse momenta $p_{t,b} \gg m_b$ or large $Q^2 \gg (2 m_b)^2$, 
the massive scheme becomes unreliable due to large terms in the 
perturbation series of the form 
$\alpha_s \ln(p^2_{t,b}/m^2_b)$ or $\alpha_s \ln(Q^2/m^2_b)$.
In this kinematic range, the massless scheme~\cite{massless} can be used, 
in which charm and beauty are treated as active flavours in both 
the proton and the photon, in addition 
to $u$, $d$ and $s$.
In this scheme, so-called excitation processes occur 
in which the beauty quark is a constituent of the resolved photon
(sketched in figures~\ref{fig:feynman}c and d) or of the proton. 

In this analysis the measurements are compared with NLO calculations
in the massive scheme for both photoproduction~\cite{Frixione:1995qc}
and DIS~\cite{Harris:1995dv}.
NLO calculations in the massless scheme 
are not yet available for the exclusive final state considered in this measurement.
The data are also compared with the predictions of the Monte Carlo 
simulations PYTHIA~\cite{PYTHIA}, RAPGAP~\cite{Jung:1993gf}
and CASCADE~\cite{Jung:2000hk}.
In the Monte Carlo simulations leading order matrix elements are implemented
and higher orders are approximated using parton showers radiated from the 
initial and final state partons.
PYTHIA and RAPGAP use the DGLAP~\cite{Gribov:ri}
parton evolution equations, while
CASCADE contains an implementation of the CCFM~\cite{ccfm} evolution 
equation. In CASCADE the direct process $\gamma g \rightarrow 
b\bar{b}$ is implemented using off-shell matrix elements convoluted 
with $k_t$-unintegrated parton distributions in the proton.

\section{Detector Description}
\label{sec:det}

The H1 detector is described in detail in \cite{Abt:1997xv} and
only the components most relevant for this analysis 
are briefly discussed here.
A right handed coordinate system is employed that has its $z$-axis 
pointing in the proton beam, or forward, direction and $x$ ($y$) pointing
in the horizontal (vertical) direction.
Charged particles 
are measured in the Central Tracking Detector (CTD) which covers the
range in pseudo-rapidity between 
$-1.74 < \eta < 1.74$\footnote{The pseudo-rapidity 
$\eta$ corresponding to a polar angle $\theta$ is given by
$\eta = -\ln \; \tan (\theta/2)$.}.
The CTD comprises two large
cylindrical Central Jet Chambers (CJCs) and two $z$-chambers 
arranged concentrically around the beam-line 
within a solenoidal magnetic field of 1.15 T. 
The CTD also provides trigger
information which is based on track segments in the $r$-$\phi$ plane 
measured in the CJCs and the $z$-position of the interaction 
vertex obtained from a double 
layer of multi-wire proportional chambers.

The CTD tracks are linked with hits in the Central
Silicon Track detector (CST)~\cite{Pitzl:2000wz}, which consists of two 
36 cm long concentric cylindrical
layers of silicon strip detectors, 
surrounding the beam pipe  
at radii of $57.5$ mm and $97$ mm from the beam axis. 
The CST covers a pseudo-rapidity range of  $-1.3< \eta < 1.3\;$
for tracks passing through both layers.
The double-sided silicon detectors 
provide resolutions of 12 $\mu$m in $r$-$\phi$ and 25 $\mu$m in $z$.
Average hit efficiencies are 97\% (92\%) in 
$r$-$\phi$ ($z$).
For a CTD track with CST 
$r$-$\phi$ hits in both layers,
the transverse distance of closest approach $dca$ of the track 
to the nominal vertex in $x$-$y$ can be measured with a resolution of 
$\sigma_{dca} \approx 33\;\mu\mbox{m} 
\oplus 90 \;\mu\mbox{m} /p_t [\mbox{GeV}]$, where the first
term  represents the intrinsic resolution (including alignment
uncertainties) and the second term is 
the contribution
from multiple scattering in the beam pipe and the CST;  
$p_t$ is the transverse momentum of the track.

The track detectors are surrounded in the forward and
central directions, $-1.5 < \eta < 3.4$, by a fine grained Liquid Argon 
calorimeter (LAr) and in the backward region,
\mbox{$-4.0 < \eta < -1.4$}, by a lead-scintillating fibre calorimeter
SpaCal~\cite{Appuhn:1996na} with electromagnetic and hadronic sections.
%
The SpaCal is used primarily in this analysis to
detect the scattered electron in DIS events and to select
photoproduction events, in which case the scattered electron 
is not detected.
The calorimeters are surrounded by the solenoidal 
magnet and the iron return yoke which is instrumented with
16 layers of limited streamer tubes in the range $-2.5 < \eta < 3.4$.
In the central pseudo-rapidity range, studied in this paper, 
the instrumented iron allows high efficiency detection of muon 
tracks with $p_t^\mu \gtrsim 2$~GeV.

The $ep$ luminosity is determined by measuring the QED
bremsstrahlung $(ep \to ep\gamma)$ event rate by detecting the
radiated photon in a calorimeter located at $z = -103$\ m.
%
\section{Event Selection and Reconstruction}
\label{sec:sel}

The data were recorded in the years 1999 and 2000 and 
correspond to an integrated luminosity of \mbox{$\sim 50$~pb$^{-1}$}.
During this time HERA was operated with positrons 
of 27.6 GeV and protons of 920~GeV energy. 
The events were triggered by requiring  that there be
signals from the central drift chambers and the multi-wire
proportional chambers in coincidence with
signals from the scattered positron in the backward
calorimeter (DIS sample) or with
signals from the instrumented iron (photoproduction).

At least one muon is required with a transverse momentum
$p_t^{\mu}>2.5$ GeV.
Muons are identified as track segments 
in the barrel part of the instrumented iron.
The iron track segments must be well matched to a 
track reconstructed in the CTD.
At least two CST $r$-$\phi$-hits have to be associated with the muon track
and it is required that the combined CTD-CST $r$-$\phi$-track fit 
probability exceeds 10\%.  
The muon momentum is reconstructed using the
CTD-CST track information.
The CST hit requirements for the muon track 
restrict the allowed range of $ep$ interactions
along the $z$-axis to $|z_{vtx}| \le 20\;\mbox{cm}$.

Jets are reconstructed 
using the inclusive $k_t$ algorithm~\cite{Ellis:1993tq,Catani:1993hr}
in the $p_t$ recombination scheme (see~\cite{Butterworth:2002xg}), 
giving massless jets.
The algorithm, with a distance parameter in the $\eta$-$\phi$ plane of 1, 
is applied to all hadronic final state particles, which are 
reconstructed using a combination
of tracks and calorimeter energy deposits~\cite{Adloff:1997mi}.
The muon, as measured in the CTD and CST, is one of the
particles which is input to the jet algorithm.
The muon track is required to be associated 
with one of the selected jets by the jet algorithm.

Photoproduction events are selected by demanding that there be no 
electromagnetic cluster in the backward and central calorimeter with an energy
of more than 8 GeV. 
This cut restricts the accepted range of negative
four-momentum transfer squared to $Q^2<1$ GeV$^2$ with a mean of about $0.07$ GeV$^2$.  
For the photoproduction sample, a cut on the
inelasticity \mbox{0.2 $<y<$ 0.8} is applied, where $y$ is reconstructed 
using the relation $y=\sum_{h}(E-p_z)/2 E_e$~\cite{JB}.  
Here, $E$ and $p_z$ are the energies and $z$-components of the
momenta of the hadronic final state particles, $h$,
and $E_e$ is the positron beam energy. 
The final photoproduction event sample consists of 1745 events. 
The number of events containing more than one muon candidate 
is less than 1\%. 
The jet algorithm is applied in the
laboratory frame and 
at least two jets are required with 
transverse momenta \mbox{$p_t^{jet_{1(2)}}>7(6)$ GeV} in
the pseudo-rapidity range $|\eta^{jet}|<2.5$.
The fraction of the photon energy entering
the hard interaction is estimated using the 
observable
$$
\xgobsm = \frac{ \sum_{Jet_1}(E-p_z) + 
                 \sum_{Jet_2}(E-p_z)}{\sum_{h}(E-p_z)},
$$
where the sums in the numerator run over
the particles associated with the two jets
and that in the denominator over all 
detected hadronic final state particles.
For the direct process (figure~\ref{fig:feynman}a),
$\xgobsm$ approaches unity, as the hadronic final state consists of
only the two hard jets and the proton remnant in the
forward region which contributes little to $\sum_{h}(E-p_z)$.
In resolved processes $\xgobsm$ can be small.

DIS events are selected by requiring a scattered positron signal 
with an energy of at least 8 GeV in the SpaCal.
To suppress photoproduction background and to reduce the fraction of
events with significant initial state QED radiation,
events are rejected if \mbox{$\sum (E-p_z)<45$ GeV}. 
Here, $E-p_z$ is summed over all final state particles
including the scattered positron.
The kinematic
variables $Q^2$ and $y$ are reconstructed using
the $e\Sigma$ method \cite{Bassler:1994uq},
which combines the hadronic final state and the
positron measurements.
%
The scaling variable $x$ is subsequently calculated
using the relation $x=Q^2/ys$, where $s$ is the $ep$ centre-of-mass energy
squared. 
Events are selected in the range
$2<Q^2<100$ GeV$^2$ and $0.1<y<0.7$.
The jet algorithm is applied in the
Breit frame~\cite{breit} and at least one jet
with transverse momentum $p^{Breit}_{t,jet}>6$ GeV
is required, with which the muon must be associated.
The final DIS event sample consists of 776 events.

Table~\ref{tab:cuts} summarises the selection
cuts for the two samples. The selection cuts for the
photoproduction sample are somewhat tighter, 
due to the muon trigger acceptance and to 
suppress background events.

\begin{table}[h]
\begin{center}
\begin{tabular}{|l|c|c|}
\hline & 
Photoproduction & DIS \\
 \hline
$Q^2$ [GeV$^2$] & $< 1$ &  2~...~100 \\ 
$y$  & 0.2~...~0.8 & 0.1~...~0.7 \\
\hline
Frame            & laboratory  & Breit \\
$\#$ jets           & $\ge 2$       &  $\ge 1$ \\
$p_t^{jet}$ [GeV]      & $>7(6)$    & $>6$ \\
$|\eta_{lab}^{jet}|$    & $<2.5$ & $<2.5$ \\
\hline
$\mu$ Iron link probability  & $>10\%$ & $>5\%
$ \\
$\mu$ CST link probability  & $>10\%$ & $>10\%
$ \\
$\#$ CST hits  & $\ge 2$ & $\ge 2$ \\
$\eta^{\mu}$    & $-0.55~...~1.1$ & $-0.75~...~1.15$ \\
$p_t^{\mu}$ [GeV]    & $>2.5$ & $>2.5$ \\
\hline
$\#$ events & 1745 & 776 \\
\hline
\end{tabular}
\caption{Selection cuts for the photoproduction and DIS data samples 
and the number of selected events.}
\label{tab:cuts}
\end{center}
\end{table}

%
%
\section{Monte Carlo Simulations}
\label{sec:mc}
The Monte Carlo generators 
PYTHIA~\cite{PYTHIA} 
and RAPGAP~\cite{Jung:1993gf}
are used for the description of the signal and background 
distributions, the determination of efficiencies and acceptances
and for systematic studies.
The track resolutions were adjusted to describe the data.
In addition, systematic cross checks and estimates of model 
dependence are performed using the CASCADE
generator~\cite{Jung:2000hk}.
The measured beauty production cross sections
are also compared
with the predictions of these three generators.
The basic parameter choices for the various 
pQCD programs are summarised in table~\ref{tab=MC}.

All three Monte Carlo generators are used to produce
large samples of beauty and charm events with decays into muons
$ep\ra e\bbb X \rightarrow \mu X'$ or
$ep\ra e\ccb X \rightarrow \mu X'$. The Peterson
fragmentation function~\cite{Peterson:1982ak} is used for the hadronisation 
of the heavy quarks.
For systematic cross checks, samples using the
Lund string fragmentation model~\cite{Andersson:1983ia} are generated.
Each of these Monte Carlo samples corresponds
to at least forty times the luminosity of the data.
In addition, PYTHIA and RAPGAP 
event samples for light quark ($uds$), $c$ and $b$ events 
without muon requirements
are generated with six times the luminosity of the data. 
The latter samples are used for the simulation
of the background due to hadrons 
misidentified as muons and decays of light hadrons into muons.
All generated events are passed through a detailed
simulation of the detector response based on the
GEANT program \cite{geant} and reconstructed using the 
same reconstruction software as used for the data.

For the  measurements in photoproduction, PYTHIA is used 
in an inclusive mode in which direct and resolved 
events are generated using massless matrix elements for all 
quark flavours (MSTP(14)=30~\cite{PYTHIA}).
Beauty (charm) events are separated from light quark events by requiring
that there be at least one beauty (charm) quark in the list of outgoing
hard partons.
Approximately 35\% of the PYTHIA beauty cross section in the
measured range is due to
resolved photon processes and these are dominated
by the flavour excitation component.
For the measurements in the DIS region, RAPGAP is used to generate 
the direct production process 
in the massive mode (IPRO=14~\cite{Jung:1993gf}).
RAPGAP is interfaced with the program HERACLES~\cite{Kwiatkowski:1990es} which
simulates QED initial and final state radiation.
Additional event samples are generated using 
the Monte Carlo generator CASCADE~\cite{Jung:2000hk}.
Comparisons of the measurements with the predictions
of CASCADE are made using version 1.2 of the program
and the J2003 parton density 
distributions~\cite{Jung:2003wu,Andersen:2003xj,Hansson:2003xz}.

All generators use 
the JETSET part of the PYTHIA~\cite{PYTHIA} program 
to simulate the hadronisation and decay processes.
The branching ratios for the direct semileptonic decays 
$b\rightarrow \mu X$ and for the indirect decays into muons
via charm, anticharm, $\tau$ and $J/\Psi$ decays
are in agreement with the world average 
values \cite{Hagiwara:fs}. 
The total branching ratio for beauty decays into muons
is 21\% \cite{Hagiwara:fs}.
The decay lifetimes of the beauty and charm hadrons
are set to the values reported in \cite{Hagiwara:fs}. 
The muon momentum spectrum in the
rest frame of the decaying $b$-flavoured hadrons, 
as modeled by JETSET, is in agreement with the spectrum 
measured at $e^{+}e^{-}$ colliders~\cite{Abe:2002du,Aubert:2002uf}.
%
%
%

\begin{table}[t]
\begin{center}
{\footnotesize
\begin{tabular}{l@{\hspace{.4cm}}c@{\hspace{.4cm}}c@{\hspace{.4cm}}c@{\hspace{.4cm}}c@{\hspace{.4cm}}c@{\hspace{.4cm}}}
  & {\bf PYTHIA} 
  & {\bf RAPGAP}
  & {\bf CASCADE} 
  & {\bf FMNR}  
  & {\bf HVQDIS} 
  \rule[-2mm]{0mm}{5mm}  
  \\
\hline
\hline
Version       &   6.1   
              &   2.8  
              &   1.00/09;\/ 1.2
              &         
              &   1.4 
              \rule[-1mm]{0mm}{5mm} 
              \\
\hline
Proton PDF   
              &  CTEQ5L~\cite{Lai:1999wy} 
              &  CTEQ5L
              &  JS2001~\cite{Jung:2000hk}
              &  CTEQ5M~\cite{Lai:1999wy} 
              &  CTEQ5F4~\cite{Lai:1999wy} 
              \rule[-1mm]{0mm}{5mm}
              \\

              &   {}
              &   {}
              &  J2003~\cite{Jung:2003wu,Andersen:2003xj,Hansson:2003xz}
              &   {}
              &   {} 
              \\
Photon PDF   
              &  GRV-G LO~\cite{Gluck:1991}
              &  {}
              &  {}        
              &  GRV-G HO~\cite{Gluck:1991}
              &  {}
              \rule[-2mm]{0mm}{5mm}
              \\
\hline
$\Lambda_{QCD}^{(4)}  \ [ {\rm GeV} ]$  & 0.192 & 0.192 & 0.2 & 0.326 & 0.309 \\

\hline
Renorm. scale  $\mu_r^2$ \rule[-1mm]{0mm}{6mm} 
               &  $m_q^2 + p_{tq\bar{q}}^2$
               &  $Q^2 + p_{tq\bar{q}}^2$
               &  $\hat{s} +  p_{tq\bar{q}}^2$
               &  $m_b^2 + p_{tb\bar{b}}^2$
               &  $m_b^2 + p_{tb\bar{b}}^2$
               \\[1mm]
Factor. scale  $\mu_f^2$ \rule[-3mm]{0mm}{6mm}
               &  $m_q^2 + p_{tq\bar{q}}^2$
               &  $Q^2 + p_{tq\bar{q}}^2$
               &  $\hat{s} + Q_t^2$
               &  $m_b^2 + p_{tb\bar{b}}^2$
               &  $m_b^2 + p_{tb\bar{b}}^2$
               \rule[-2mm]{0mm}{5mm}  
               \\
\hline
$m_b \ [ {\rm GeV} ]$           \rule[-1mm]{0mm}{6mm}
               & 4.75  
               & 4.75 
               & 4.75 
               & 4.75 
               & 4.75 
               \\
$m_c \ [ {\rm GeV} ]$         
               & 1.5  
               & 1.5  
               & 1.5  
               & {} 
               & {} 
               \rule[-2mm]{0mm}{5mm}  
               \\  
\hline 
Peterson $\,\epsilon_b$  \rule[-1mm]{0mm}{6mm}
               & 0.0069
               & 0.0069
               & 0.0069
               & 0.0033
               & 0.0033
               \\ 
Peterson $\,\epsilon_c$  
               & 0.058
               & 0.058
               & 0.058
               & { }
               & { }
               \\ 
\hline
\hline
\end{tabular}
\caption{Parameters used in the leading order Monte Carlo simulations and the 
NLO programs.
Here $\mu_r$ and $\mu_f$ denote the renormalisation and factorisation
scales, 
$m_q$ the heavy quark masses,
$p_{tq\bar{q}}$  the average of the transverse momenta
of the two heavy quarks,
$\hat{s}$ and  $Q_t^2$ 
the heavy quark system centre-of-mass energy squared and
transverse momentum squared, respectively, and 
$\epsilon_q$ the Peterson fragmentation parameters.}
\label{tab=MC}}
\end{center}
\end{table}

%
\section{NLO QCD Calculations}
\label{sec:nlo}

The NLO pQCD calculations are performed 
in the massive scheme using the program FMNR~\cite{Frixione:1995qc}
in the photoproduction regime and 
the program HVQDIS~\cite{Harris:1995dv}
for the DIS case.
Both programs provide weighted parton level events 
with two or three outgoing partons, i.e.\,a $b$ quark, 
a $\bar{b}$ quark and possibly an additional light parton.
The calculations are performed in the $\overline{\rm MS}$-scheme
using the parameters given in table~\ref{tab=MC}.

The $b$ quark is `hadronised'
into a $b$-flavoured hadron by rescaling the
three-momentum of the quark using
the Peterson fragmentation function~\cite{Peterson:1982ak} 
with the parameter $\epsilon_b = 0.0033$~\cite{Nason:1999zj}.
The programs are extended to include the decay of the $b$-flavoured hadron 
into a final state with a muon. 
The muon decay spectrum is taken from JETSET \cite{PYTHIA} 
and includes direct and indirect decays of 
$b$-flavoured hadrons into muons. 
Parton level jets are 
reconstructed by applying the $k_t$ jet algorithm to
the outgoing partons.

Corrections to the hadron level are calculated using 
the PYTHIA and RAPGAP Monte Carlo event generators.
PYTHIA and RAPGAP parton level jets are reconstructed 
from the generated quarks and gluons after the 
parton showering step.
At the hadron level, jets are reconstructed by applying
the jet algorithm to all final state particles,
after the decay of the beauty or charm hadrons.
The jet and the muon selection cuts are 
applied to the generator samples.
In each kinematic bin of the measurement,
the ratio of the PYTHIA or RAPGAP hadron level and parton level 
cross sections is calculated and applied as a correction factor 
to the NLO calculation.
The parton to hadron level corrections range typically 
from $-30\%$ to $+5\%$ in both photoproduction and DIS.
The corrections are negative at small muon and jet transverse momenta
and positive at the largest transverse momenta. 
The corrections obtained using CASCADE are consistent
with the values from PYTHIA (photoproduction) and RAPGAP (DIS).

The theoretical uncertainties of the NLO calculations
are estimated in the following way:
The $b$ quark mass and the renormalisation and factorisation
scales are varied simultaneously
from $m_b = 4.5$ GeV and $\mu_r = \mu_f = m_T/2$ \/ to \/ 
$m_b = 5$ GeV and $\mu_r = \mu_f = 2m_T$, where
$m_T=\sqrt{m_b^2 + p_{tb\bar{b}}^2}\;$, and $p_{tb\bar{b}}$
is the average of the transverse momenta of the two $b$ quarks.
This leads to a maximum change of the cross section of typically
25\% in photoproduction (FMNR) and 15--20\%
in DIS (HVQDIS).
The cross section variation when using
other proton structure functions such as MRSG or 
MRST1 \cite{Martin}
is less than 8\% in all regions of the measurement.
The uncertainty due to variations of the fragmentation parameter 
$\epsilon_b$ by $25\%$ is below 3\%.
These cross section variations are added in quadrature
to estimate the total systematic uncertainty of the NLO predictions
for each bin of the measurement.
%
\section{Comparison of the Data with Monte Carlo Simulations}
\label{sec:control}
Detailed comparisons are performed of the data with the
Monte Carlo simulations.
The good agreement of the PYTHIA simulation with the photoproduction data 
is illustrated in figure~\ref{fig:final-control}.
The distributions of the muon transverse momentum \ptmu, the
pseudo-rapidity \etamu, the jet transverse momentum
$p_t^{jet_{1(2)}}$ and \xgobs are shown. 
The data are compared with the sum
of the contributions from beauty, charm and light quark events, 
the relative fractions of which are taken from the 
two-dimensional fit discussed in section~\ref{sec:fit}.
The number of events in the simulation is normalised to that of the data.
It is observed that the shapes of the distributions 
are rather similar for the 
beauty, charm and light quark events
except for the muon transverse momentum, where the 
spectrum is harder for beauty than for the other components.
In figure~\ref{fig:dfinal-control}, the
distributions for the DIS sample are shown. 
The photon virtuality $Q^2$, 
the inelasticity $y$, Bjorken-$x$, 
the muon transverse momentum \ptmu\ and
the transverse momentum 
$p^{Breit}_{t,jet}$ of the selected jet in the Breit frame 
are well described by the RAPGAP Monte Carlo simulation.
The CASCADE Monte Carlo simulation also provides a 
good description of the data in both 
photoproduction and DIS (not shown). 
%
\section{Determination of Beauty and Background Contributions}
\label{sec:fit}
The signed impact parameter \del of the muon track 
and the transverse momentum \ptrel of the muon track 
relative to the axis of the associated jet are used
to determine the fraction of beauty events in the data.
For each muon candidate,  \del is calculated in the plane transverse 
to the beam axis. 
The magnitude of \del is given by the $dca$ of the track
to the primary vertex. 
The sign is defined as positive if the angle between the
jet axis and the line joining the primary vertex
to the point of closest approach of the track
is less than $90^{\circ}$, and is defined as negative otherwise.
Figures~\ref{fig:final-signal}a and~\ref{fig:dfinal-signal}a
show the distributions of \del for the photoproduction and DIS samples, 
respectively.
The decay of long-lived particles mainly leads to positive 
impact parameters, whereas particles produced at the vertex
yield a symmetric distribution centered at zero with finite width 
due to the track and primary vertex resolutions.

The transverse beam interaction region at HERA, 
in the following termed the `beam spot', 
has an approximately Gaussian profile.
For the data period studied in this paper the beam spot size
is determined to be about 145 $\mu$m  in the horizontal
and 25 $\mu$m in the vertical direction.
For each event the $ep$ collision point is determined 
in a primary vertex fit using the weighted
average of the beam spot position, taking the above widths
as errors, and position information from selected tracks in the event.
The muon track under consideration is excluded from the fit.
An average muon impact pa\-ra\-me\-ter resolution
of 80 $\mu$m is achieved, with comparable contributions
from the muon track resolution and the uncertainty in the
primary vertex position.

The transverse momentum \ptrel of the muon track 
(figures~\ref{fig:final-signal}b and~\ref{fig:dfinal-signal}b)
is calculated relative to the direction of the associated jet
according to the formula
\[ p_t^{rel} = 
\frac{| 
\vec{p_{\mu}} 
   \times 
( \vec{p}_{jet} - \vec{p_{\mu}} ) |
}{
| \vec{p}_{jet} - \vec{p_{\mu}} | }.  
\]
The quantities  $\vec{p_{\mu}}$ and $\vec{p}_{jet}$ 
are the momentum vectors of the muon and the jet in the laboratory frame,
respectively.

The fraction of muons in the data that originate from beauty events 
is determined using a likelihood fit to the two-dimensional distribution of
$\delta$ and $p_t^{rel}$ in the range \mbox{$-0.1<\delta<0.1$ cm} and
$0 <  p_t^{rel} < 3.6 $ GeV.
The combination of the two independent observables 
\del and \ptrel in the fit results in a significant improvement
in the statistical precision of the measurement
and a reduced sensitivity of the measured $b$ fraction 
to systematic uncertainties in the modelling of charm and light quark
contributions.
The shapes of the distributions \del and \ptrel for beauty\footnote{Both 
direct decays $b\rightarrow \mu X$ and indirect 
decays, e.g.\,$b\rightarrow c X'\rightarrow \mu X$,
are taken into account.}, charm and light quark events are 
obtained from the Monte Carlo simulation.  
In the fit, the relative fractions of the three 
components are adjusted such that the likelihood is maximised.
The normalisation of the sum of the three components
is fixed to match the data.
The results of the fits are illustrated in 
figures~\ref{fig:final-signal} and \ref{fig:dfinal-signal}. 
The contributions from beauty, charm and light quark events, 
with respective fractions of typically 30\%, 50\% and 20\%, are indicated.
In all bins of the measurement, the 
data are well described by the sum of the three contributions.
At large positive values of \del and at large values of \ptrelm ,  
the beauty component (dashed line) becomes dominant.

Variations of the fit procedure are investigated and
are found to give consistent results.
For example, the \del and \ptrel distributions are fitted independently of 
each other. In another fit
the relative contributions of charm and light quark events are fixed 
to the predictions of the Monte Carlo simulation.
Furthermore, the fits to the 
two-dimensional data distributions of $\delta$ and $p_t^{rel}$ are investigated
in a beauty enriched subset of the $\delta$ and $p_t^{rel}$ phase space, 
as illustrated in the figures~\ref{fig:final-highp} and~\ref{fig:dfinal-highp}.
Here, the distributions of the impact parameter \del (figures~\ref{fig:final-highp}a 
and~\ref{fig:dfinal-highp}a) and the relative transverse muon momentum 
$p_t^{rel}$
(figures~\ref{fig:final-highp}b and~\ref{fig:dfinal-highp}b) are shown 
for the cuts {\mbox{\ptrel$>1.2$ GeV} and \mbox{$\delta>0.01$~cm}, respectively. 
The lines show the predictions for the different contributions 
in the restricted samples when using the results of the fits 
to the complete samples.
The expected enhancement of the beauty contribution is observed
and the quality of the description of the data
illustrates the consistency between the results
obtained using the two observables independently.

In each kinematic bin of the measurement, 
the fit to the two-dimensional data distribution of
$\delta$ and $p_t^{rel}$
is performed,
using the data and Monte Carlo 
samples in that bin.
The fitted number of muons coming from beauty events
is translated into a cross section by correcting
for detector efficiencies, acceptances and radiative effects 
and by dividing by the integrated luminosity.
%
%
\section{Systematic Uncertainties}
\label{sec:syserr}

Systematic uncertainties of the cross section measurement are evaluated
by variations applied to the Monte Carlo simulations.
The dominant errors come from the
muon identification and muon track linking efficiencies, 
the modelling of the resolution of the muon impact parameter
and the fragmentation models.
The systematic errors assigned to the
measured cross sections are
listed in table~\ref{tab:syserr}.
%

The muon track reconstruction efficiency 
in the CTD is known to a precision of about 2\%.
An additional uncertainty of 2\% comes from the 
requirement that two CST hits be associated with the central track, 
yielding a total uncertainty for the track reconstruction efficiency of 3\%.
The uncertainty of the muon identification efficiency, including the
reconstruction in the instrumented iron and the linking with
the central track, is about 5\%.

The systematic error arising from the uncertainties of
the CTD and CST track resolutions 
is estimated by varying the muon impact parameter 
resolution in the Monte Carlo simulations by 10\%. 
This leads to cross section changes of $ 7\%$.
To substantiate this result the following cross checks are
performed and found to be consistent.
The core and the tails of the distribution of the 
impact parameter resolution are varied separately. 
The description of the beam spot ellipse
is tested by determining the cross sections
separately for two independent samples with
either more horizontal or more vertical muons.
In addition, the muon impact parameter is calculated 
with respect to the average $ep$ collision point 
instead of the primary vertex.
%

The reconstruction of the direction of the jet associated to the muon
is studied by varying the resolutions of the 
reconstructed jet directions in the Monte-Carlo 
simulations. The effect on the measured cross sections 
is about 2\%.
The jet energy scale uncertainties are estimated
by varying the LAr energy scale in the Monte Carlo simulations
by $4\%$. This leads to cross section changes of up to 4\%.

The trigger efficiencies are $ 70 \pm 3 \%$ 
for the photoproduction sample and $85 \pm 3\%$ for the DIS sample,
respectively.
For the DIS sample, the uncertainty associated with
reconstruction and identification of the scattered
positron is estimated to be less than 2\%.
The luminosity measurement contributes a global 1.5\% error.

The dependence on the physics model used for the beauty signal   
and the charm background is studied
using the CASCADE Monte Carlo generator 
instead of PYTHIA or RAPGAP, leading to 
cross section variations of about 5\%. 
Using the Lund \cite{Andersson:1983ia}
fragmentation model instead of the Peterson 
fragmentation function \cite{Peterson:1982ak}
causes changes in the measured cross sections of up to 7\%.

The modelling of the decays of the $b$-flavoured hadrons
has been tested by varying the lifetimes and branching ratios
of the different hadrons within the uncertainties
of the world average values. 
The effects on the measured cross section are at the  2\% level.
Muons from 
$\pi^{\pm}$ or $K^{\pm}$ decays
within the beam pipe and inside the sensitive volume of
the CTD and CST exhibit a broad \del distribution. 
The contribution from these events 
in the light quark Monte Carlo simulation is
varied by a factor of two, leading to 
cross section changes of 2\%.

The above systematic studies are performed separately for each 
bin of the cross section measurement.
The systematic errors   
are found to be of similar size for all bins.
For each bin a total systematic uncertainty of 14\% 
is estimated by adding all contributions to the
systematic error in quadrature.

\begin{table}[h]
\renewcommand{\arraystretch}{1.1}
\begin{center}
\begin{tabular}{|l|c|c|}
\hline
\multicolumn{1}{|c|}{Source} & 
\begin{tabular}{c}
Photoproduction \\
$\Delta \sigma / \sigma ~[\%] $ \\
\end{tabular}
&
\begin{tabular}{c}
DIS \\
$\Delta \sigma / \sigma ~[\%] $ \\
\end{tabular}
\\
\hline
\hline
 Detector efficiencies &  & \\
\bul  Scattered positron  & --   & 2 \\
\bul  Trigger efficiency  & 4   & 3 \\
\bul  Muon identification & 5   & 5 \\
\bul  CST+CTD tracks      & 3   & 3 \\
\bul  Luminosity          & 1.5 & 1.5 \\
\hline
Track reconstruction & &  \\
\bul  $\delta$ resolution & 7   & 7 \\
\hline
Jet reconstruction &  & \\
\bul  Jet axis & 2  & 2 \\
\bul  Hadronic energy scale & 4  & 4 \\
\hline
 MC model uncertainties: &  &\\
\bul PYTHIA vs.\,CASCADE   & 5  & -- \\
\bul RAPGAP vs.\,CASCADE   & --    & 5   \\
\bul Fragmentation (Peterson vs.\,Lund) & 7  & 7 \\
\bul Fragm.\,fractions, BRs, lifetimes  & 2  & 2 \\
\bul $K,\pi$ decays & 2  & 2  \\
\hline
\hline
Total &  14  & 14  \\
\hline
\end{tabular}
\caption{List of systematic uncertainties as discussed in section~\ref{sec:syserr}.
The total systematic error is obtained by adding all contributions in quadrature.}
\label{tab:syserr}
\end{center}
\end{table}
%
%
\section{Results}
\label{sec:results}
Differential beauty production cross sections are determined separately
for the photoproduction and electroproduction samples.
The results are listed in 
tables~\ref{tab:gp-xsecs},~\ref{tab:dis-xsecs} and~\ref{tab:theory} 
and displayed in figures~\ref{fig:xsec-etaptmu} to~\ref{fig:summary}.

\subsection{Photoproduction Measurement}
The visible range for the 
measurement for beauty photoproduction in dijet muon events
is \mbox{$Q^2<1$ GeV$^2$}, $0.2<y<0.8$, \ptmu\,\,$>2.5$ GeV, $-0.55<\eta^{\mu}<1.1$,
$p_t^{jet_{1(2)}} > 7(6)$ GeV and $|\eta^{jet}|<2.5$.
For this range the total cross section is measured to be
$$
 \sigma_{vis}(ep \rightarrow e b\bar{b} X 
 \rightarrow e jj\mu X') = 38.4 \pm 3.4 (stat.) \pm 5.4 (sys.)~{\rm pb}.
$$
The NLO QCD calculation 
performed in the massive scheme
with the FMNR program~\cite{Frixione:1995qc}, as described in 
section~\ref{sec:nlo},
%
%
yields for the same kinematic range
a value of $23.8 ^{+7.4}_{-5.1} {\rm pb}$ which 
is 1.5 standard deviations below the data.
The Monte Carlo programs PYTHIA and CASCADE also predict a lower
cross section than that measured in the data 
(see table~\ref{tab:theory}).
The results of all three calculations are in good agreement with each other.
An analysis using an independent H1 data sample was performed 
in~\cite{kroseberg} giving results consistent with this measurement.

Differential cross sections for beauty production are measured
as a function of several kinematic variables, shown in figure~\ref{fig:xsec-etaptmu} 
and listed in table~\ref{tab:gp-xsecs}.
The bins in which the measurement is made
are identical to the bins in which the
theory curves are presented.
The measured cross sections are quoted at the point
in the bin at which the bin-averaged cross section 
equals the differential cross section, according to the Monte Carlo
simulation. 
The data are compared with the expectations of 
the FMNR NLO QCD calculation
and the PYTHIA and CASCADE generators. 

The differential cross section measured
as a function of the muon pseudo-rapidity  \etamu\ 
(figure~\ref{fig:xsec-etaptmu}a) 
is flat in the  phase space covered.
The NLO QCD calculation describes the shape well.
This is also true for both PYTHIA and CASCADE.
The measurement agrees well with the values obtained
by the ZEUS experiment~\cite{Chekanov:2004xy}  
in their two central muon pseudo-rapidity bins, which cover
a similar phase space.
  
The differential cross sections measured
as a function of the muon transverse momentum \ptmu\ and 
of the transverse momentum of the leading jet $p_t^{jet}$  
(figure~\ref{fig:xsec-etaptmu}b and c)
fall steeply with increasing transverse momentum.
The NLO calculation clearly predicts a less steep
behaviour and is lower than the data in the lowest momentum bin
by roughly a factor of 2.5. At higher transverse momenta
better agreement is observed.
Similar conclusions can be drawn for both the 
PYTHIA and CASCADE predictions, although the latter
predicts a slightly harder $p_t^{jet}$
spectrum than the other calculations.

Figure~\ref{fig:xsec-etaptmu}d
shows the differential cross sections
as a function of $\xgobsm$.
A significant fraction of the data is found at
\xgobs$<0.75$, i.e.\,in the region in which resolved 
photon processes (figures~\ref{fig:feynman}b to d)
are enhanced.
In this observable the NLO calculation suffers 
from large uncertainties due to the scale variations.
Furthermore, the parton to hadron level 
corrections are large due to the fact that a 
single parton can produce more than one jet 
at the hadron level leading to migrations in $\xgobsm$.
Within the large uncertainties, the NLO calculation
describes the  \xgobs differential cross sections
reasonably well.
The PYTHIA simulation includes a 35\% contribution from
resolved photon processes, which are dominated by
flavour excitation processes such as those shown in  
figures~\ref{fig:feynman}c and d.
Due to the large fraction of resolved photon processes,
PYTHIA predicts a relatively high cross section value 
in the lowest \xgobs bin  
(figure~\ref{fig:xsec-etaptmu}d),
which matches the data quite well. 
However, in the largest bin, \xgobs$>0.75$,
the PYTHIA prediction is too low. 
In contrast, CASCADE succeeds in describing the
cross section of the data at large values of \xgobs
while it is too low at smaller values of $\xgobsm$.

The results of this analysis are compared
with the previous H1 measurement~\cite{Adloff:1999nr}
in which somewhat softer jet and muon cuts than in this 
analysis were used for the event selection.
The measured cross section for the process
$ep \rightarrow e b\bar{b}X \rightarrow e jj\mu X'$ 
is extrapolated to the inclusive $b$ quark cross section, 
$ep \rightarrow b\bar{b}X \rightarrow \mu X'$,
in the kinematic region $Q^2<1$ GeV$^2$, $0.1<y<0.8$, 
$p_t^{\mu}>2$~GeV and $35^{\circ}<\theta^{\mu}<130^{\circ}$, as in~\cite{Adloff:1999nr}.
The extrapolation is performed using the Monte Carlo program 
AROMA\cite{Ingelman:1996mv} which was also used 
in~\cite{Adloff:1999nr}.
The result, scaled to 820 GeV proton beam energy,
is $107.3 \pm 9.5 (stat.) \pm 15.1 (sys.)$ 
pb, which is $2.3$ standard deviations lower than the value
of $176 \pm 16 (stat.)~^{+ 26}_{-17}(sys.)$ pb
obtained in~\cite{Adloff:1999nr}.

\subsection{Electroproduction Measurement}
The beauty electroproduction cross section is measured in the visible range
\mbox{$2<Q^2<100$~GeV$^2$}, $0.1<y<0.7$, \ptmu\,\,$>2.5$ GeV, $-0.75<\eta^{\mu}<1.15$,
$p^{Breit}_{t,jet} > 6$ GeV and $|\eta^{jet}|<2.5$, yielding
$$ 
\sigma_{vis}(ep \rightarrow e b\bar{b} X \rightarrow e j\mu X') =
 16.3 \pm 2.0 (stat.) \pm 2.3 (sys.)~{\rm pb}.
$$
The prediction of the NLO QCD calculation in the 
massive scheme using the program HVQDIS
is $9.0^{+2.6}_{-1.6}$ pb,
which is 1.8 standard deviations below the data.
The Monte Carlo programs RAPGAP and CASCADE also predict a lower
cross section than that measured in the data 
(see table~\ref{tab:theory}).

Differential cross section measurements are presented in
figures~\ref{fig:dxsec-q2x} and \ref{fig:dxsec-etaptmu} 
and in table~\ref{tab:dis-xsecs}.
The data are compared with the expectations of
the HVQDIS NLO QCD calculation
and the RAPGAP and CASCADE generators. 
The differential cross section
as a function of the photon virtuality $Q^2$   
is shown in figure~\ref{fig:dxsec-q2x}a. 
The NLO calculation describes the shape well, but lies below
the data. The prediction of
CASCADE is similar to that of the NLO calculation while
RAPGAP, which only contains the direct photon contribution 
to the cross section, is somewhat further below the data.
The differential cross section
as a function of the scaling variable $x$ 
is shown in figure~\ref{fig:dxsec-q2x}b. 
The various calculations also describe the shape of the 
data well, while the overall normalisation is again too low.

In figure~\ref{fig:dxsec-etaptmu}, the differential cross sections
are presented as functions of the muon and leading jet kinematics.
The differential cross section measured
in bins of the muon pseudo-rapidity \etamu\ 
(figure~\ref{fig:dxsec-etaptmu}a) exhibits 
a rise towards the forward region, which is not reproduced
by the NLO and Monte Carlo calculations.
The differential cross sections measured
as a function of the transverse momenta of the muon \ptmu\ 
and of the jet in the Breit frame
$p^{Breit}_{t,jet}$ (figures~\ref{fig:dxsec-etaptmu}b and c)
show a steep distribution, as is the case in 
photoproduction (figures~\ref{fig:xsec-etaptmu}b and c).
The shapes of the NLO QCD, RAPGAP and CASCADE predictions 
are all very similar.
As in photoproduction, the measured electroproduction cross sections
as a function of the muon and jet transverse momenta
show a steeper behaviour than the predictions of the
NLO calculations and the Monte Carlo simulations
and significantly exceed the predictions in the lowest bins.

Figure~\ref{fig:summary} presents a summary of recent
HERA beauty cross section measurements as a function of the 
photon virtuality $Q^2$. 
The figure shows the ratios of the measured cross
sections~\cite{Aktas:2004az,Chekanov:2004xy,Chekanov:2004tk} 
and the corresponding next-to-leading order predictions where 
FMNR is used for the photoproduction and HVQDIS for the DIS region.
The dotted lines indicate the typical theory error due to scale
uncertainties. 
General agreement is seen between the results from H1 and ZEUS,
the data tending to be somewhat above the 
NLO predictions.
%
\section{Conclusions}
\label{sec:conclusions}
Differential beauty production cross sections are measured
in $ep$ collisions at HERA both in photoproduction ($Q^2<1$~GeV$^2$)  
and in electroproduction ($2<Q^2<100$~GeV$^2$).
The event selection requires the presence of 
at least one jet (two jets) in the DIS (photoproduction) sample 
and a muon in the central pseudo-rapidity range.
For the first time at HERA, beauty events are identified using
both the transverse momentum of the muon relative to the jet axis and 
the large impact parameter of the muon.
The cross sections presented here are in general agreement with those 
obtained by the ZEUS experiment.
The data are compared with
predictions based on NLO QCD calculations 
in the massive scheme and with the expectations of
Monte Carlo generators which use leading order 
matrix elements and parton showers.
The predictions from all these calculations are 
similar in both normalisation and shape.

In both photoproduction and DIS, the total cross section measurements 
are somewhat higher than the predictions.
The excess is observed mainly at small muon and jet transverse momenta, 
while at larger momenta a reasonable description is obtained.
In photoproduction a significant contribution to the
cross section is observed in the region of small values of
the  observable $\xgobsm$, where  
contributions from resolved photon events are enhanced.
In this region the best description of the data is given by the PYTHIA simulation, 
which incorporates flavour excitation processes in which
the beauty quark is a constituent of the resolved photon or the proton.
In DIS, the observed excess is pronounced at large muon pseudo-rapidities.
The shape of the cross section as function of the photon 
virtuality $Q^2$ is reproduced by the QCD calculations over the
full range covered by the measurement presented in this paper, 
from quasi-real photons up to virtualities of about $4m_b^2$.

%
\section*{Acknowledgements}

We are grateful to the HERA machine group whose outstanding
efforts have made this experiment possible. 
We thank
the engineers and technicians for their work in constructing and
maintaining the H1 detector, our funding agencies for 
financial support, the
DESY technical staff for continual assistance
and the DESY directorate for support and for the
hospitality which they extend to the non-DESY 
members of the collaboration.

\clearpage

\begin{table}
\begin{center}
\begin{tabular}{|cc|c|c|c|c|c|}
 \hline
 \multicolumn{3}{|c|}{ } & Measurement & \multicolumn{3}{|c|}{Experimental Errors} \\ \hline \hline
 \multicolumn{2}{|c|}{$\eta^{\mu}$-range} & $\eta^{\mu}$ & $d\sigma/d\eta^{\mu}$ & stat. & syst. & total \\
 \multicolumn{2}{|c|}{ } & &  [pb]& [pb] & [pb] & [pb] \\
 \hline
           --0.55 &             --0.15 &             --0.35 &   19.1 &    3.4 &    2.7 &    4.3 \\
           --0.15 &   \phantom{--}0.25 &   \phantom{--}0.05 &   23.4 &    4.0 &    3.3 &    5.2 \\
 \phantom{--}0.25 &   \phantom{--}0.65 &   \phantom{--}0.45 &   23.9 &    3.9 &    3.3 &    5.1 \\
 \phantom{--}0.65 &   \phantom{--}1.10 &   \phantom{--}0.85 &   21.8 &    3.8 &    3.0 &    4.9 \\
 \hline
 \hline
 \multicolumn{2}{|c|}{$p_t^{\mu}$-range} & $p_t^{\mu}$ & $d\sigma/dp_t^{\mu}$ & stat.    & syst.    & total \\
 \multicolumn{2}{|c|}{  [GeV]}         &  [GeV]       &  [pb/GeV]            & [pb/GeV] & [pb/GeV] & [pb/GeV] \\
 \hline
 2.5 & \phantom{1}3.3 & 2.9 & 24.4\phantom{0} & 3.3\phantom{0} & 3.4\phantom{0} & 4.8\phantom{0} \\
 3.3 & \phantom{1}5.0 & 4.1 &  8.1\phantom{0} & 1.1\phantom{0} & 1.1\phantom{0} & 1.6\phantom{0} \\
 5.0 &           12.0 & 7.2 &  1.15 &      0.18 &           0.16 &           0.24 \\
 \hline
 \hline
 \multicolumn{2}{|c|}{$p^{jet}_{t}$-range} & $p^{jet}_t$  & $d\sigma/dp^{jet}_t$ & stat. & syst. & total \\
 \multicolumn{2}{|c|}{ [GeV]}& [GeV] &  [pb/GeV] & [pb/GeV] & [pb/GeV] & [pb/GeV] \\
 \hline
  \phantom{1}7.0 &  10.0 &  \phantom{1}8.8 & 6.3\phantom{0} & 0.8\phantom{0} & 0.9\phantom{0} & 1.2\phantom{0} \\
            10.0 &  14.0 &            11.7 & 2.9\phantom{0} & 0.4\phantom{0} & 0.4\phantom{0} & 0.6\phantom{0} \\
            14.0 &  25.0 &            18.3 & 0.83 &           0.14 &           0.12 &           0.18 \\
 \hline
 \hline
 \multicolumn{2}{|c|}{$\xgobsm$-range} & $\xgobsm$ & $d\sigma/d\xgobsm$ & stat. & syst. & total \\
 \multicolumn{2}{|c|}{   } &  & [pb] & [pb] & [pb] & [pb] \\
 \hline
   0.20 &   0.50 &   0.35 &   17.2 &    4.5 &    \phantom{1}2.4 &    \phantom{1}5.1 \\
   0.50 &   0.75 &   0.63 &   21.4 &    5.2 &    \phantom{1}3.0 &    \phantom{1}6.0 \\
   0.75 &   1.00 &   0.88 &   86.6 &    9.1 &              12.1 &              15.2 \\
 \hline
\end{tabular}
\end{center}
\caption{Differential cross sections for the process
$ep \rightarrow e b \bar{b} X \rightarrow e jj\mu X'$ 
for the photoproduction sample in the kinematic range
$Q^2<1$ GeV$^2$, $0.2<y<0.8$, \ptmu $>2.5$ GeV, $-0.55<\eta^{\mu}<1.1$,
$p_t^{jet_{1(2)}} > 7(6)$ GeV and $|\eta^{jet}|<2.5$.}
\label{tab:gp-xsecs}
\end{table}

\begin{table}
\begin{center}
\begin{tabular}{|cc|c|c|c|c|c|}
 \hline
 \multicolumn{3}{|c|}{ } & Measurement & \multicolumn{3}{|c|}{Experimental Errors} \\ \hline \hline
 \multicolumn{2}{|c|}{$Q^2$-range }&$ Q^2$ & $d\sigma/dQ^2$  & stat. & syst. & total \\
 \multicolumn{2}{|c|}{[GeV$^2$]}& [GeV$^2$] &  [pb/GeV$^2$] & [pb/GeV$^2$] & [pb/GeV$^2$] & [pb/GeV$^2$] \\
 \hline
  \phantom{1}2.0 & \phantom{00}5.0 & \phantom{0}3.5 & 1.55\phantom{0} & 0.30\phantom{0} & 0.22\phantom{0} & 0.37\phantom{0} \\
  \phantom{1}5.0 & \phantom{0}18.0 & \phantom{0}9.5 & 0.297           & 0.075 &           0.042 &           0.086 \\
 18.0 &           100.0 &           45.0 & 0.091           & 0.016 &           0.013 &           0.020 \\
 \hline
 \hline
 \multicolumn{2}{|c|}{$x$-range} &$ \log x$ & $d\sigma/d\log x$ & stat. & syst. & total \\
 \multicolumn{2}{|c|}{} & & [pb]& [pb] & [pb] & [pb] \\
 \hline
 --4.5 &  --3.8 &  --4.15 &   6.40 &   1.33 &   0.90 &   1.60 \\
 --3.8 &  --3.1 &  --3.45 &   9.72 &   1.71 &   1.36 &   2.18 \\
 --3.1 &  --2.4 &  --2.75 &   6.68 &   1.44 &   0.93 &   1.71 \\
 \hline
 \hline
 \multicolumn{2}{|c|}{$\eta^{\mu}$-range} &$\eta^{\mu}$ & $d\sigma/d\eta^{\mu}$ & stat. & syst. & total \\
 \multicolumn{2}{|c|}{} & &  [pb]&[pb] & [pb] & [pb] \\
 \hline
           --0.75 &           --0.12 &           --0.4 & \phantom{1}5.36 & 1.42 &           0.75 &           1.60 \\
           --0.12 & \phantom{--}0.50 & \phantom{--}0.2 & \phantom{1}7.40 & 1.64 &           1.03 &           1.94 \\
 \phantom{--}0.50 & \phantom{--}1.15 & \phantom{--}0.8 & 12.6\phantom{0} & 1.9\phantom{0} & 1.8\phantom{0} & 2.6\phantom{0} \\
 \hline
 \hline
 \multicolumn{2}{|c|}{$p_t^{\mu}$-range }&$ p_t^{\mu}$ & $d\sigma/dp_t^{\mu}$ & stat. & syst. & total \\
 \multicolumn{2}{|c|}{[GeV]}&[GeV] &  [pb/GeV]& [pb/GeV] & [pb/GeV] & [pb/GeV] \\
 \hline
 2.5 &   \phantom{1}3.0 & 2.8 & 11.3\phantom{00} &           2.4\phantom{00} & 1.6\phantom{00} & 2.9\phantom{00} \\
 3.0 &   \phantom{1}3.8 & 3.4 & \phantom{1}8.05\phantom{0} & 1.39\phantom{0} & 1.13\phantom{0} & 1.79\phantom{0} \\
 3.8 &             12.0 & 6.4 & \phantom{1}0.622 &           0.124           & 0.087 &           0.151 \\
 \hline
 \hline
 \multicolumn{2}{|c|}{$p^{Breit}_{t,jet}$-range}&$p^{Breit}_{t,jet}$ & $d\sigma/dp^{Breit}_{t,jet}$ & stat. & syst. & total\\
 \multicolumn{2}{|c|}{ [GeV]}&[GeV] &  [pb/GeV] & [pb/GeV] & [pb/GeV] & [pb/GeV] \\
 \hline
 \phantom{1}6.0 & \phantom{1}8.5 & \phantom{1}7.2 & 2.20\phantom{0} & 0.52\phantom{0} & 0.31\phantom{0} & 0.60\phantom{0} \\
 \phantom{1}8.5 &           12.0 & 10.0 &           1.96\phantom{0} & 0.31\phantom{0} & 0.27\phantom{0} & 0.42\phantom{0} \\
           12.0 &           30.0 & 18.5 &           0.183 &           0.043 &           0.026 &           0.050 \\
 \hline
\end{tabular}
\end{center}
\caption{Differential cross sections for the process
$ep \rightarrow e b \bar{b} X \rightarrow e j\mu X'$ 
for the electroproduction sample in the kinematic range
$2<Q^2<100$~GeV$^2$, $0.1<y<0.7$, \ptmu $>2.5$ GeV, $-0.75<\eta^{\mu}<1.15$,
$p^{Breit}_{t,jet} > 6$ GeV and 
$|\eta^{jet}|<2.5$.}
\label{tab:dis-xsecs}
\end{table}

\begin{table}
\begin{center}
\begin{tabular}{|r|c||r|c|}
\hline
\multicolumn{2}{|c||}{Photoproduction} & \multicolumn{2}{|c|}{Electroproduction} \\
\multicolumn{2}{|c||}{$\sigma(ep \rightarrow e b \bar{b} X \rightarrow e jj\mu X')$[pb]} & 
\multicolumn{2}{|c|}{$\sigma(ep \rightarrow e b \bar{b} X \rightarrow e j\mu X')$[pb]} \\
\hline
Data & $ 38.4 \pm 3.4 \pm 5.4 $ & Data  &  $16.3 \pm 2.0 \pm 2.3$ \\
FMNR & $ 23.8 ^{+7.4}_{-5.1}$ & HVQDIS & $9.0^{+2.6}_{-1.6}$  \\
PYTHIA & $20.9$ & RAPGAP     & $6.3$  \\
CASCADE & $22.6$  & CASCADE &  $9.8$ \\ \hline
\end{tabular}
\end{center}
\caption{Measured cross sections with their statistical and systematic errors 
and corresponding predictions from NLO QCD calculations and Monte Carlo simulations 
in the kinematic range $Q^2<1$ GeV$^2$, $0.2<y<0.8$, \ptmu $>2.5$ GeV, $-0.55<\eta^{\mu}<1.1$,
$p_t^{jet_{1(2)}} > 7(6)$ GeV and $|\eta^{jet}|<2.5$ (photoproduction)
 and in the kinematic range $2<Q^2<100$~GeV$^2$, $0.1<y<0.7$, \ptmu $>2.5$ GeV, $-0.75<\eta^{\mu}<1.15$,
$p^{Breit}_{t,jet} > 6$ GeV and $|\eta^{jet}|<2.5$ (electroproduction). 
The errors for the predictions from FMNR and HVQDIS 
give the systematic uncertainties as estimated from scale variations 
(see text).}
\label{tab:theory}
\end{table}

%
\begin{figure}
\setlength{\unitlength}{1cm} 
\begin{picture}(14.0,19.)
\put(6.3,18.5){\Large \sf Photoproduction}
\put(-.8,12.4){\epsfig{file=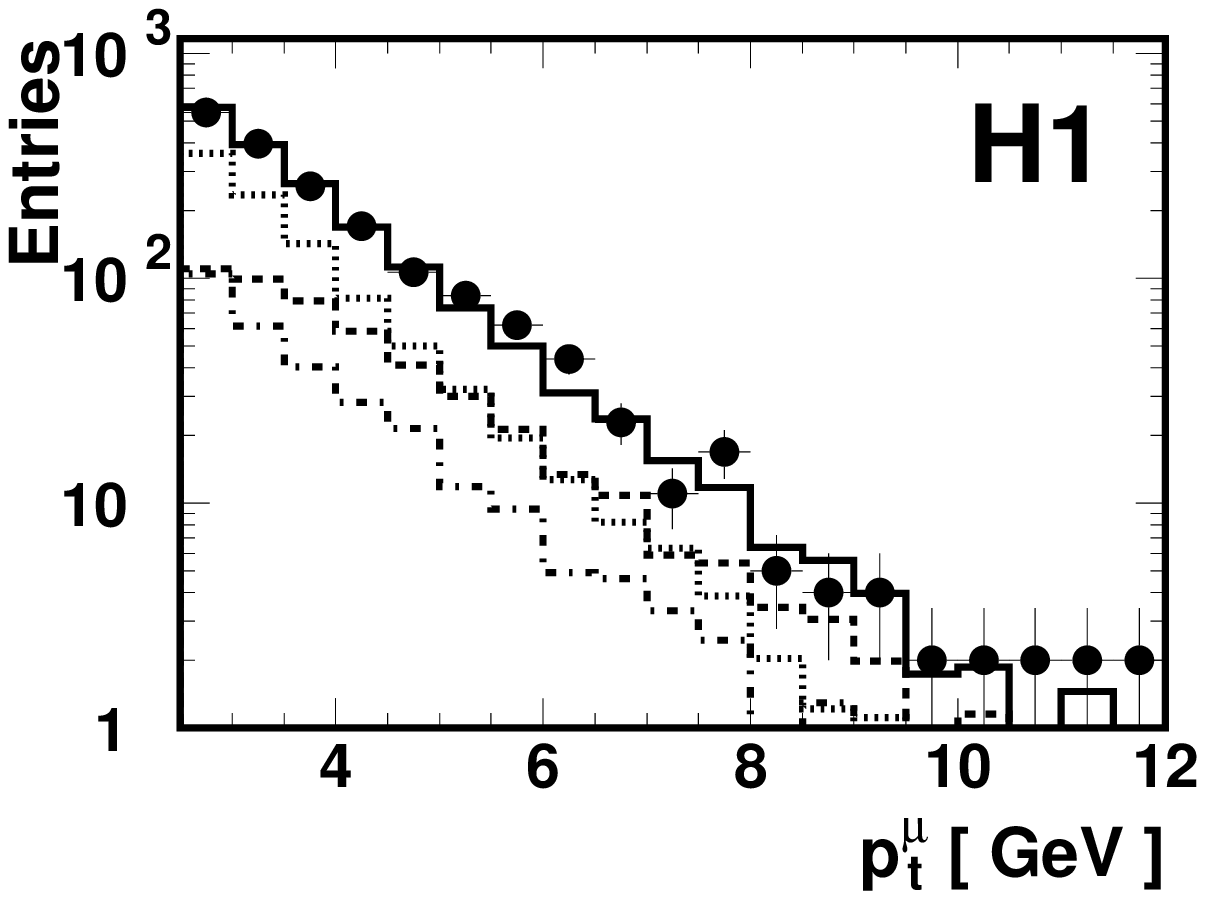,width=9cm}}
\put(7.5,12.4){\epsfig{file=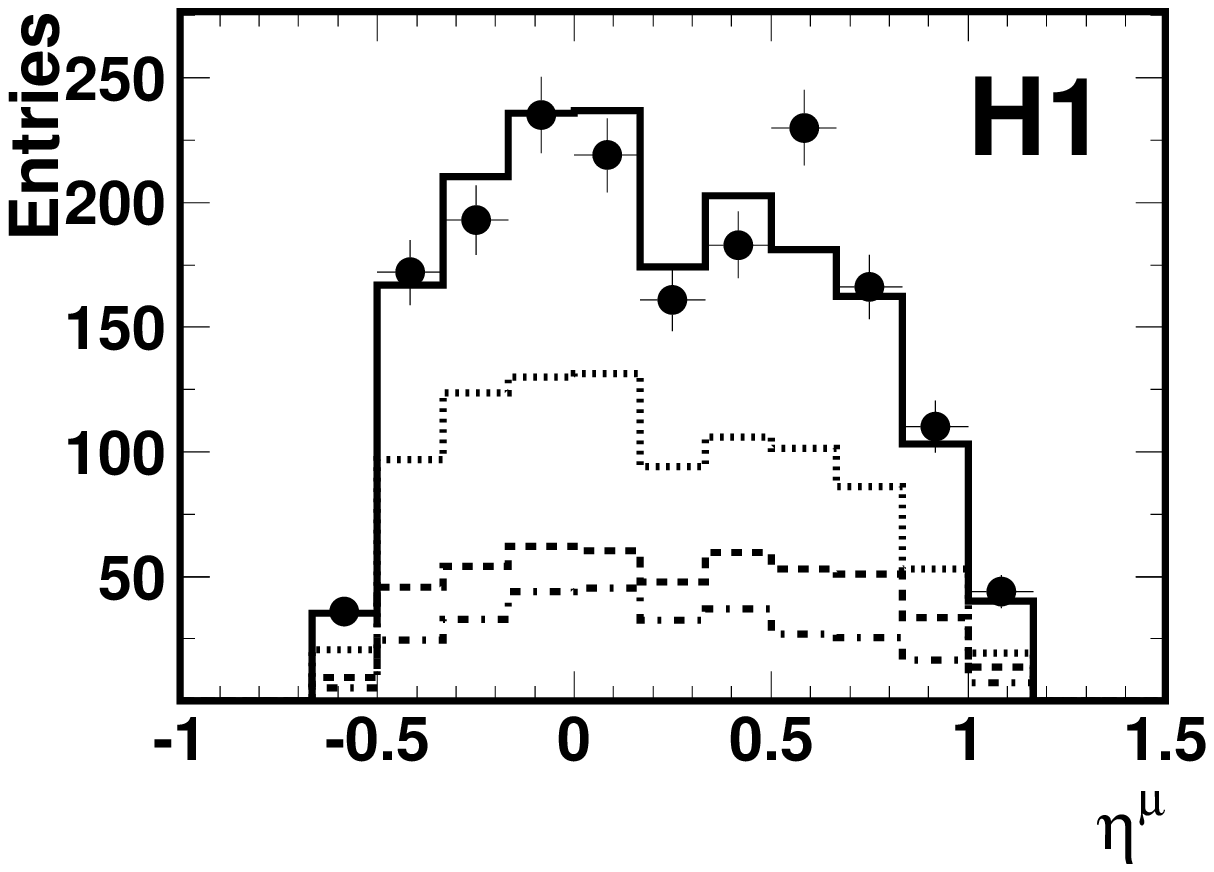,width=9cm}}
\put(-.8,6.2){\epsfig{file=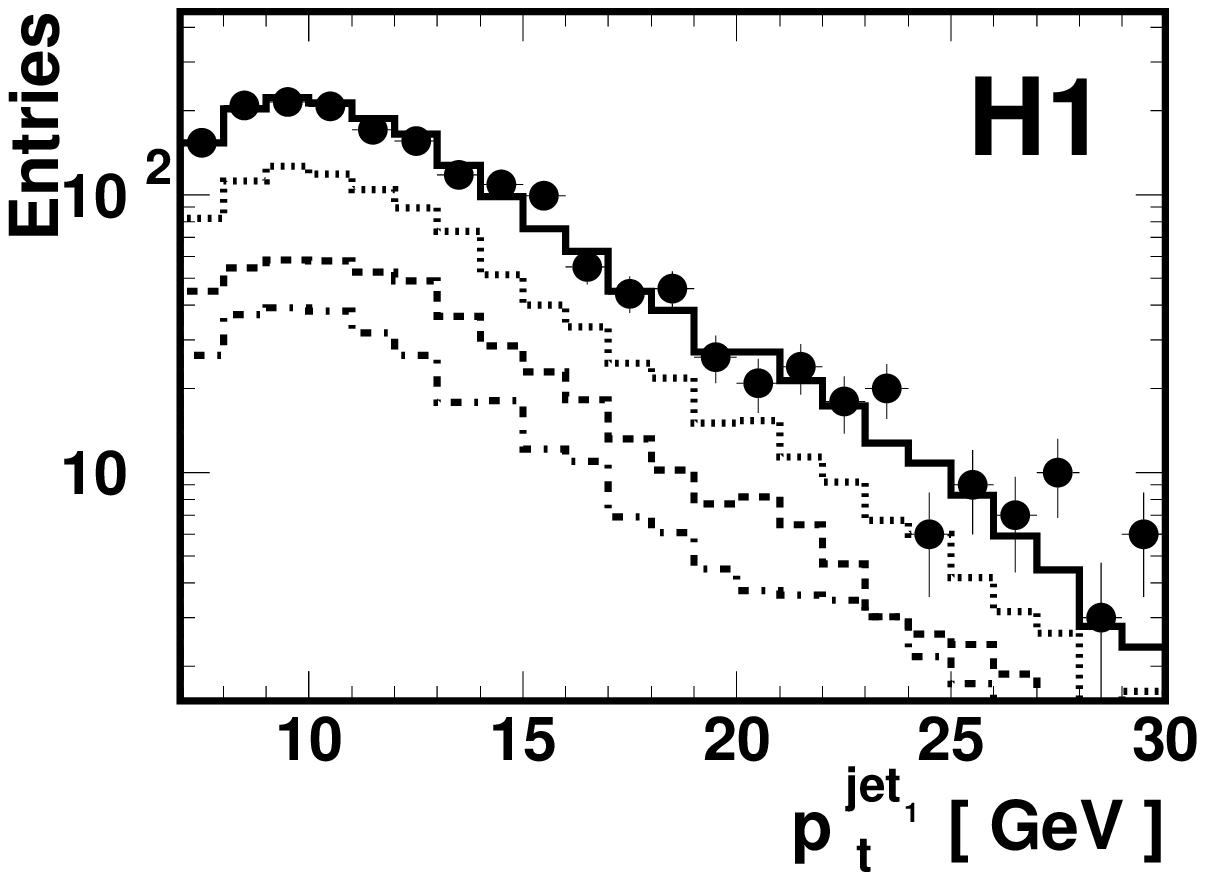,width=9cm}}
\put(7.5,6.2){\epsfig{file=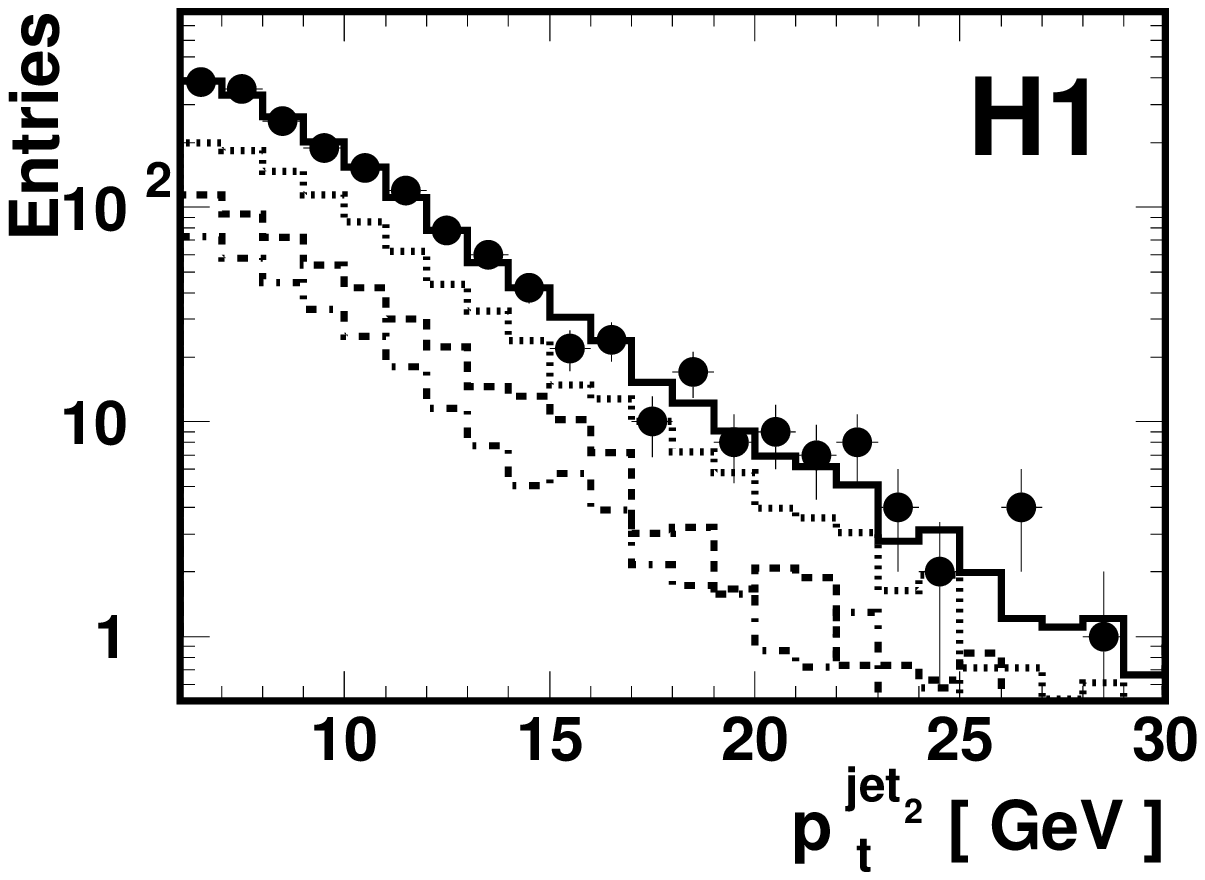,width=9cm}}
\put(-.8,0.){\epsfig{file=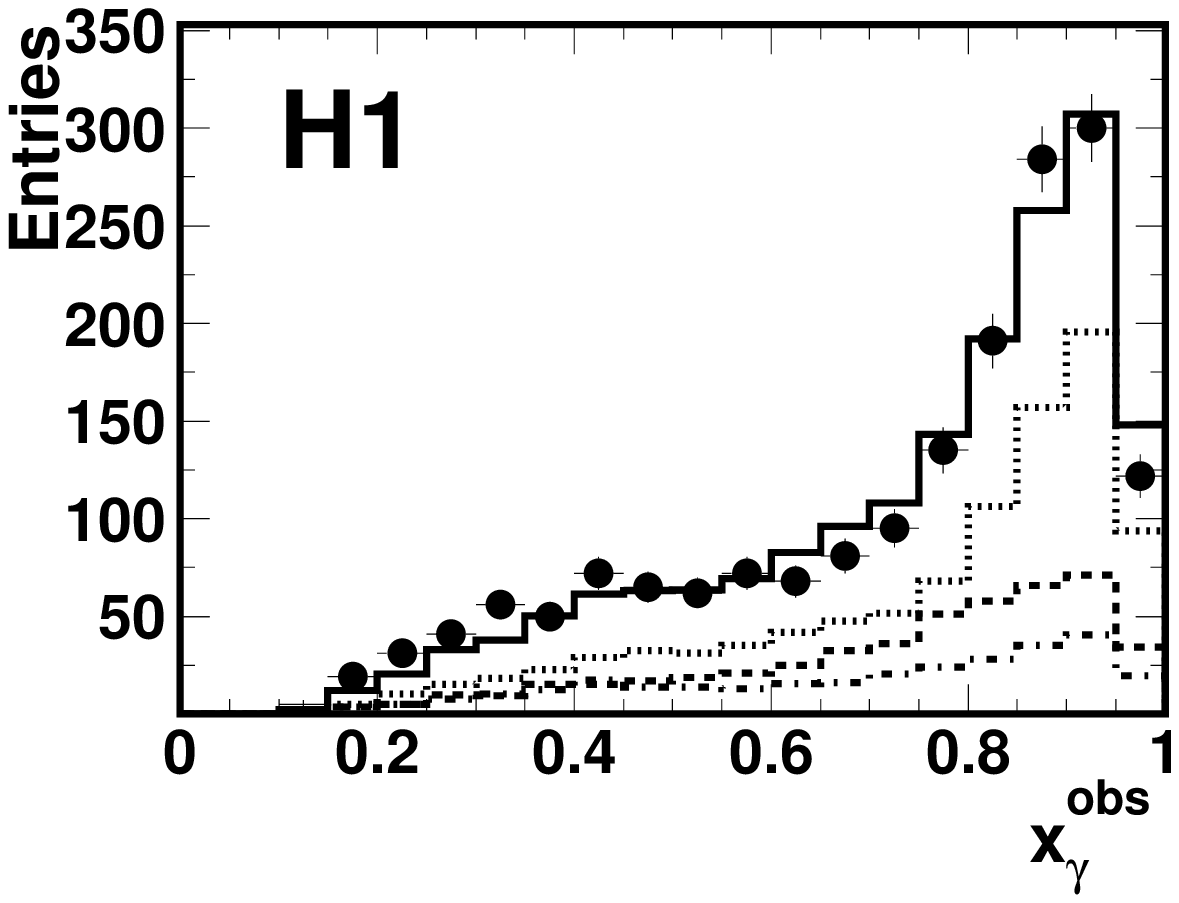,width=9cm}}
\put(8.5,0.5){\epsfig{file=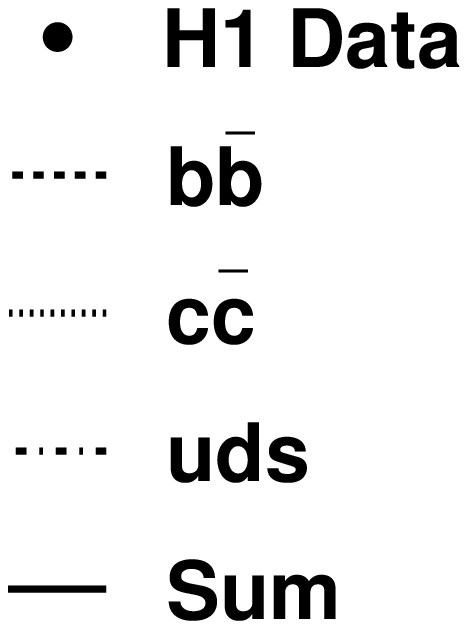,width=8cm}}
\put( 4.0,17.4){\large a)}
\put(10.0,17.4){\large b)}
\put( 4.0,11.1){\large c)}
\put(11.5,11.1){\large d)}
\put( 4.0, 4.9){\large e)}
\end{picture}
\caption{Distributions in photoproduction
of a) the muon transverse momentum \ptmu, 
b) the pseudo-rapidity of the muon \etamu, 
c) and d) the transverse momenta $p_t^{jet_{1(2)}}$ of
the highest and the second-highest $p_t$ jets, respectively 
and e) the observable $\xgobsm$.
Included in the figure are the
estimated contributions of events arising from 
$b$ quarks (dashed line), \mbox{$c$ quarks} (dotted line)
and light quarks (dash-dotted line).
The shapes of the distributions from the different sources 
are taken from the PYTHIA Monte Carlo simulation
and their relative fractions are determined from a fit to the 
two-dimensional data distribution of \ptrel and \del (see text).}
\label{fig:final-control}
\end{figure} 

\newpage

\begin{figure}
\setlength{\unitlength}{1cm} 
\begin{picture}(14.0,19.)
\put(6.3,18.5){\Large \sf Electroproduction}
\put(-.8,12.4){\epsfig{file=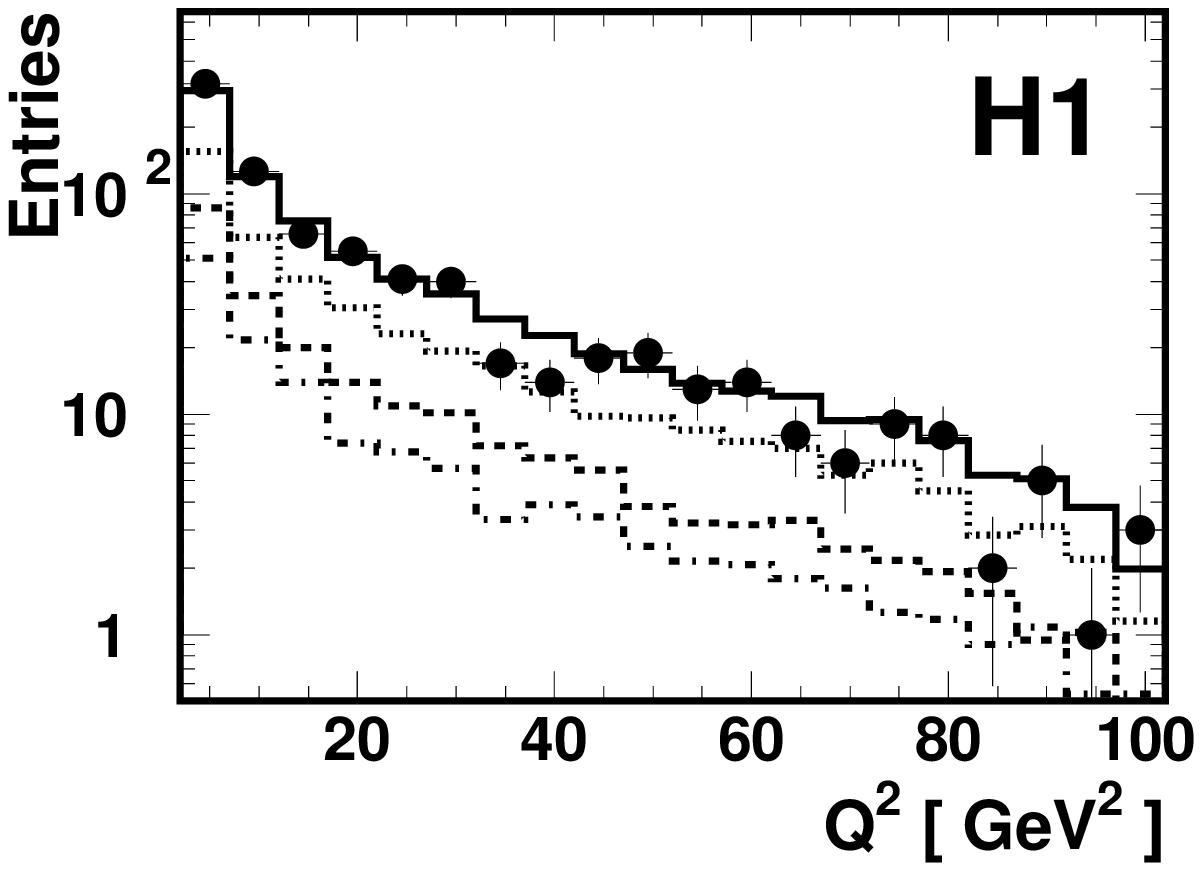,width=9cm}}
\put(7.5,12.4){\epsfig{file=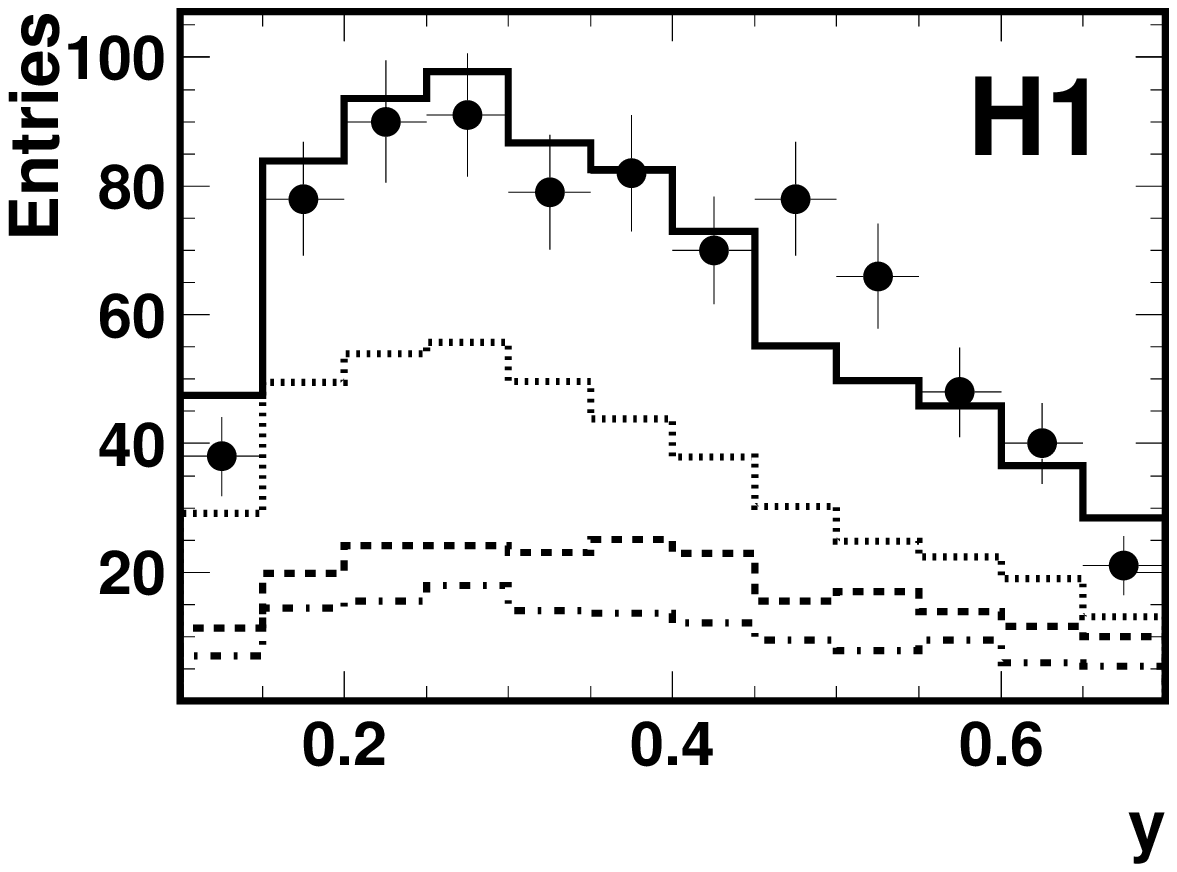,width=9cm}}
\put(-.8,6.2){\epsfig{file=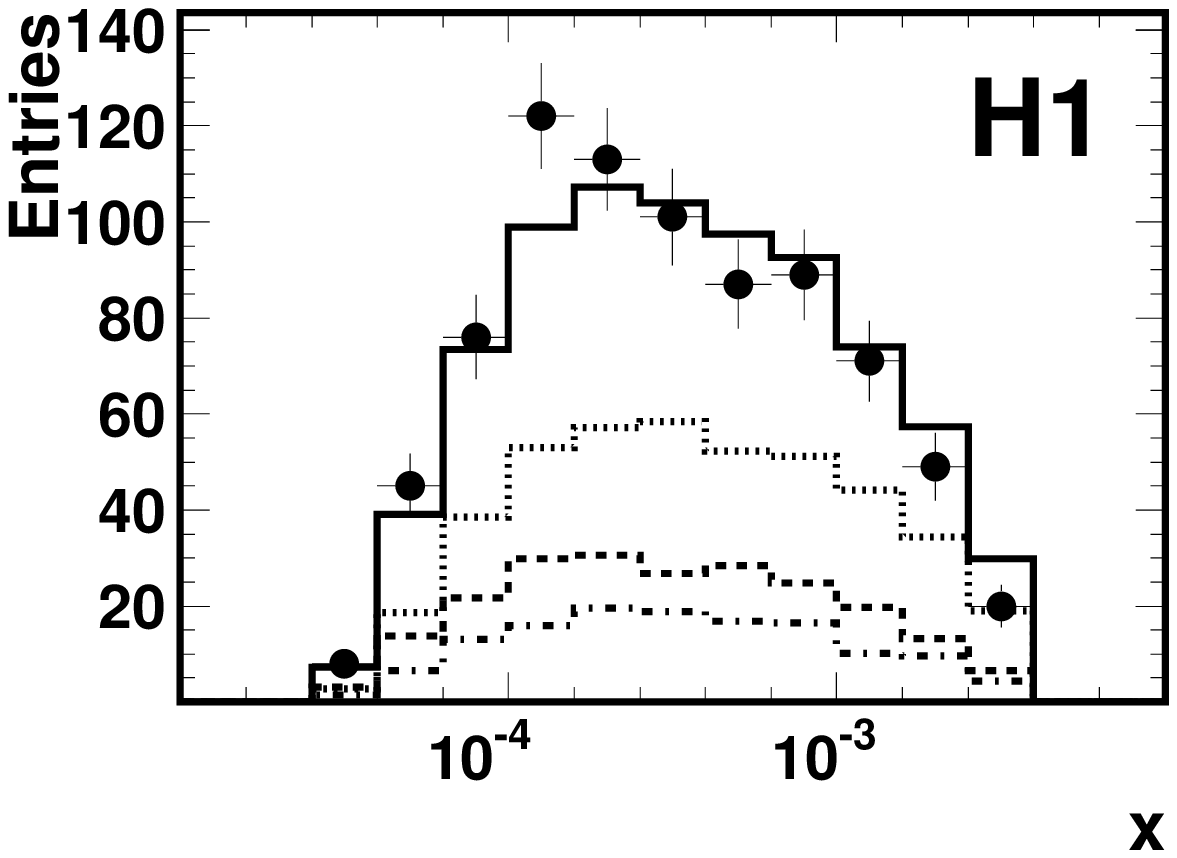,width=9cm}}
\put(7.5,6.2){\epsfig{file=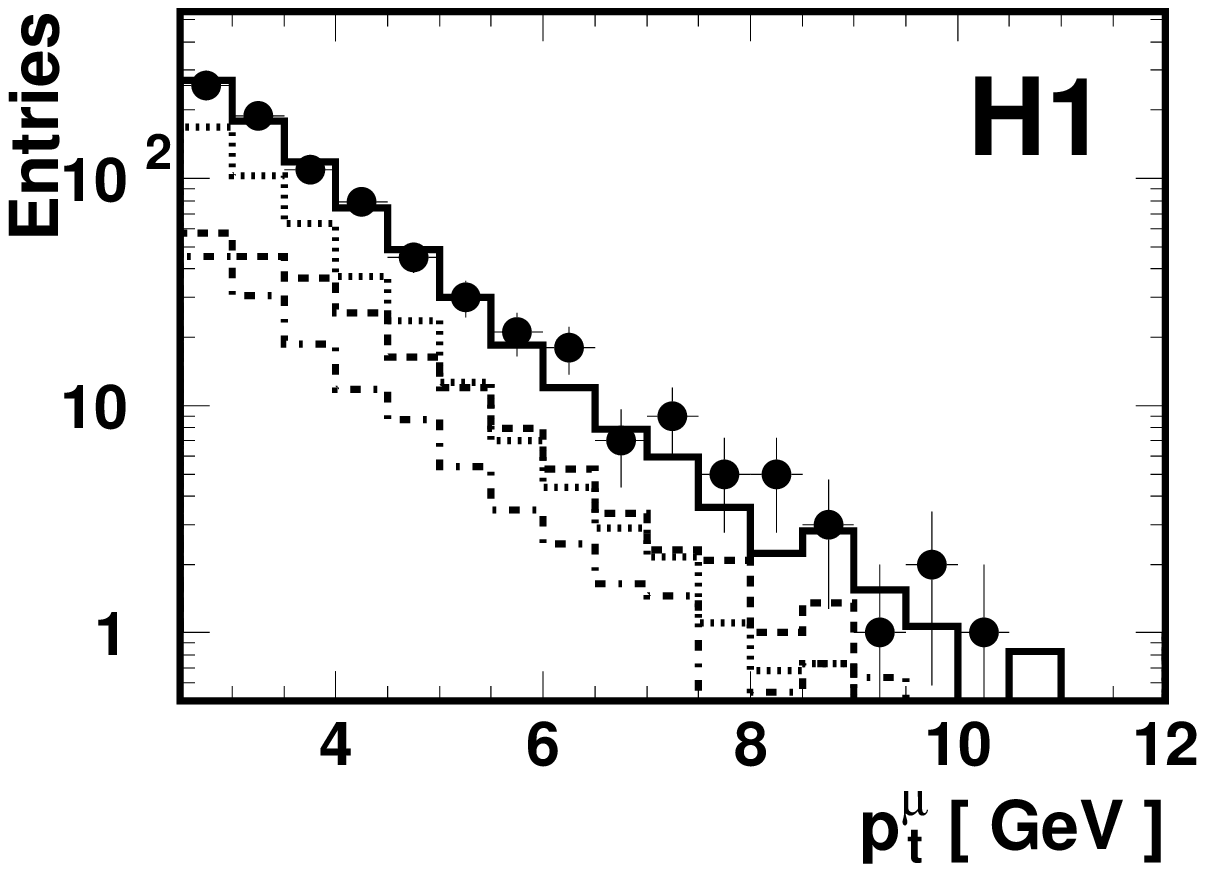,width=9cm}}
\put(-.8,0.){\epsfig{file=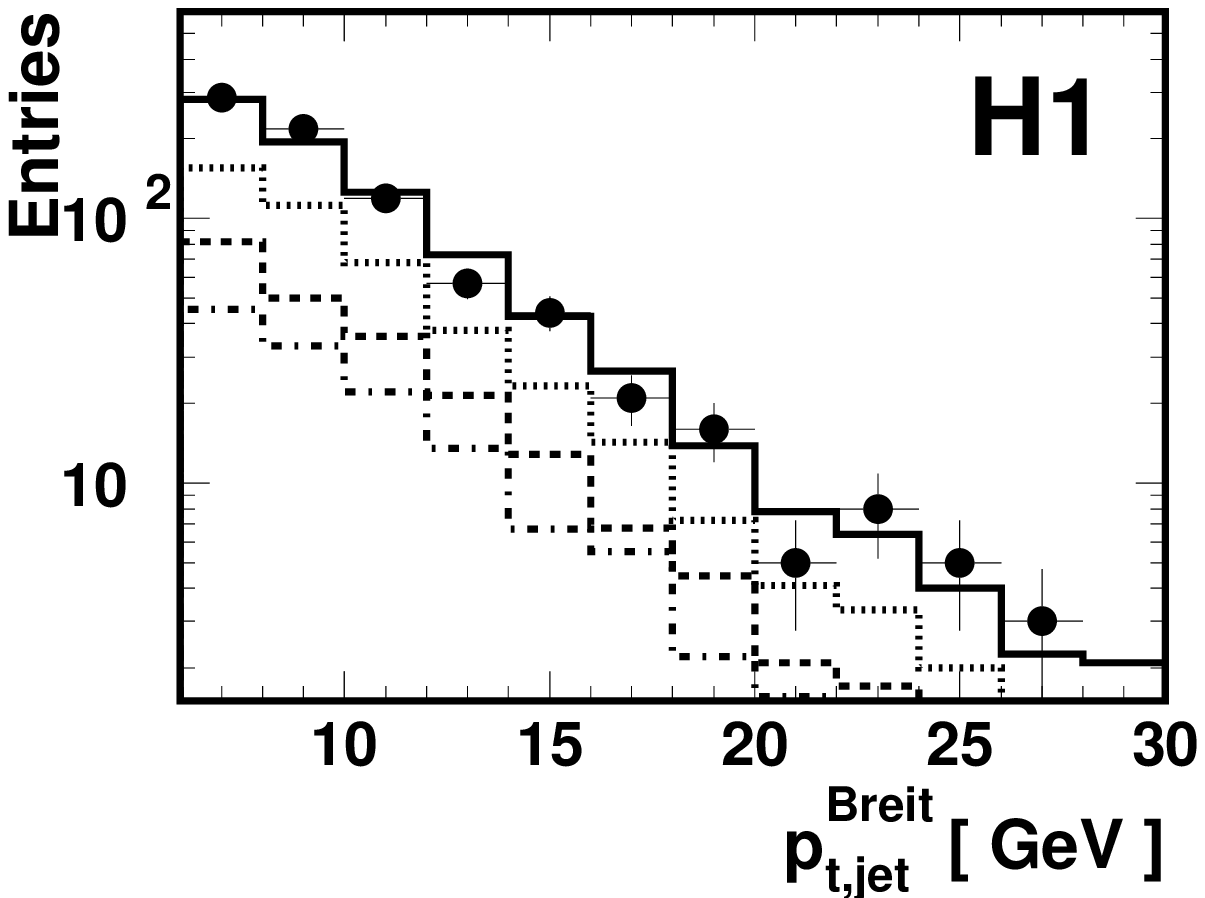,width=9cm}}
\put(8.5,0.5){\epsfig{file=d05-004f2f.eps,width=8cm}}
\put( 2.0,17.4){\large a)}
\put(13.0,17.4){\large b)}
\put( 2.0,11.1){\large c)}
\put(13.0,11.1){\large d)}
\put( 4.0, 4.9){\large e)}
\end{picture}
\caption{Distributions in electroproduction
of 
a) the  photon virtuality $Q^2$, 
b) the inelasticity $y$, 
c) Bjorken x, 
d) the muon transverse momentum and 
e) the transverse momentum $p^{Breit}_{t,jet}$ 
of the selected jet in the Breit frame.
Included in the figure are the
estimated contributions of events arising from 
$b$ quarks (dashed line), \mbox{$c$ quarks} (dotted line)
and light quarks (dash-dotted line).
The shapes of the distributions from the different sources 
are taken from the RAPGAP Monte Carlo simulation
and their relative fractions are determined from a fit to the 
two-dimensional data distribution of \ptrel and \del (see text).}
\label{fig:dfinal-control}
\end{figure}

\newpage
\begin{figure}
\setlength{\unitlength}{1cm}
\begin{picture}(14.0,9.)
\put(6.7,7.9){\large \sf Photoproduction}
\put(-.8,-.3){\epsfig{file=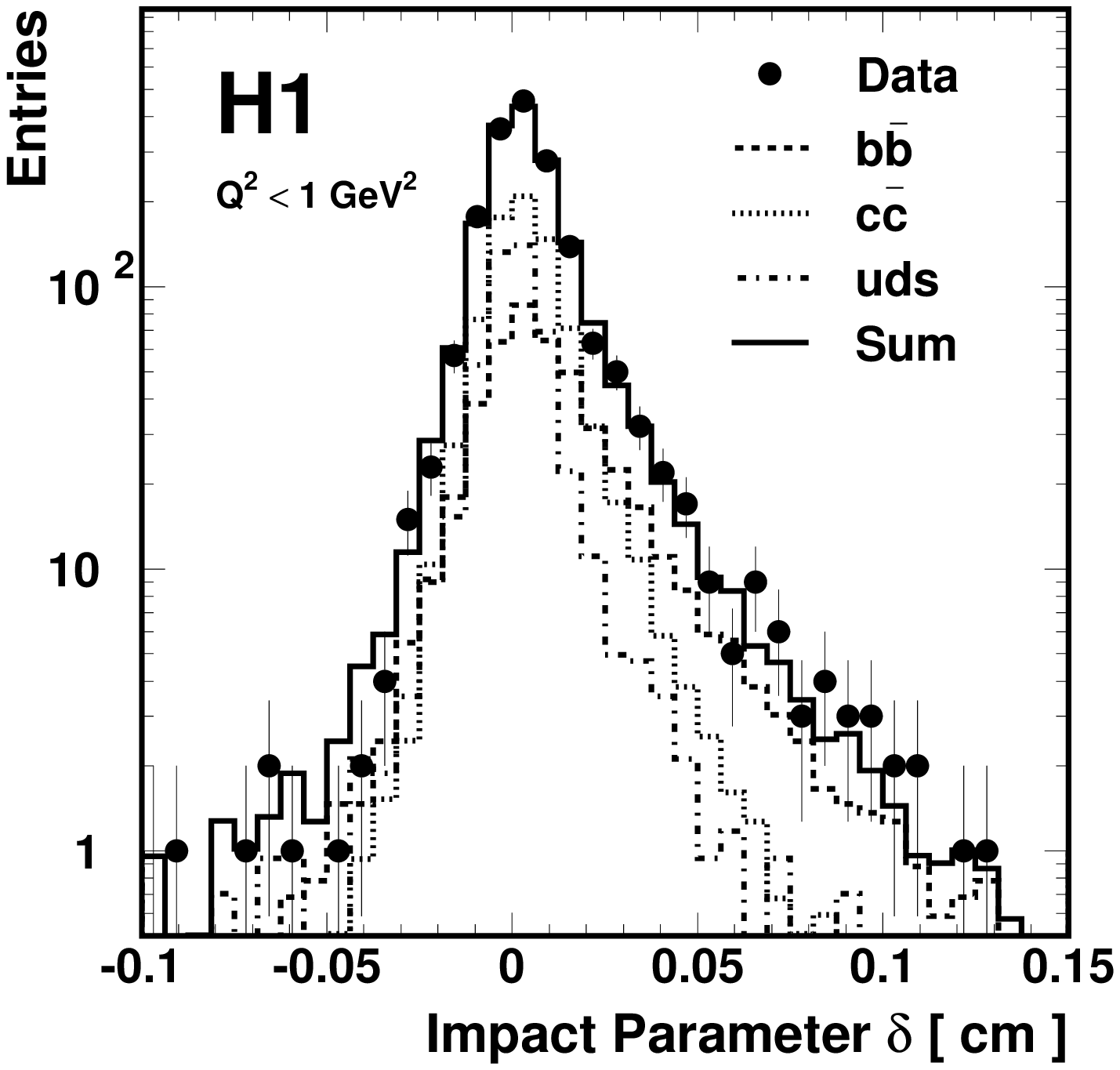,width=9cm}}
\put(7.5,-.3){\epsfig{file=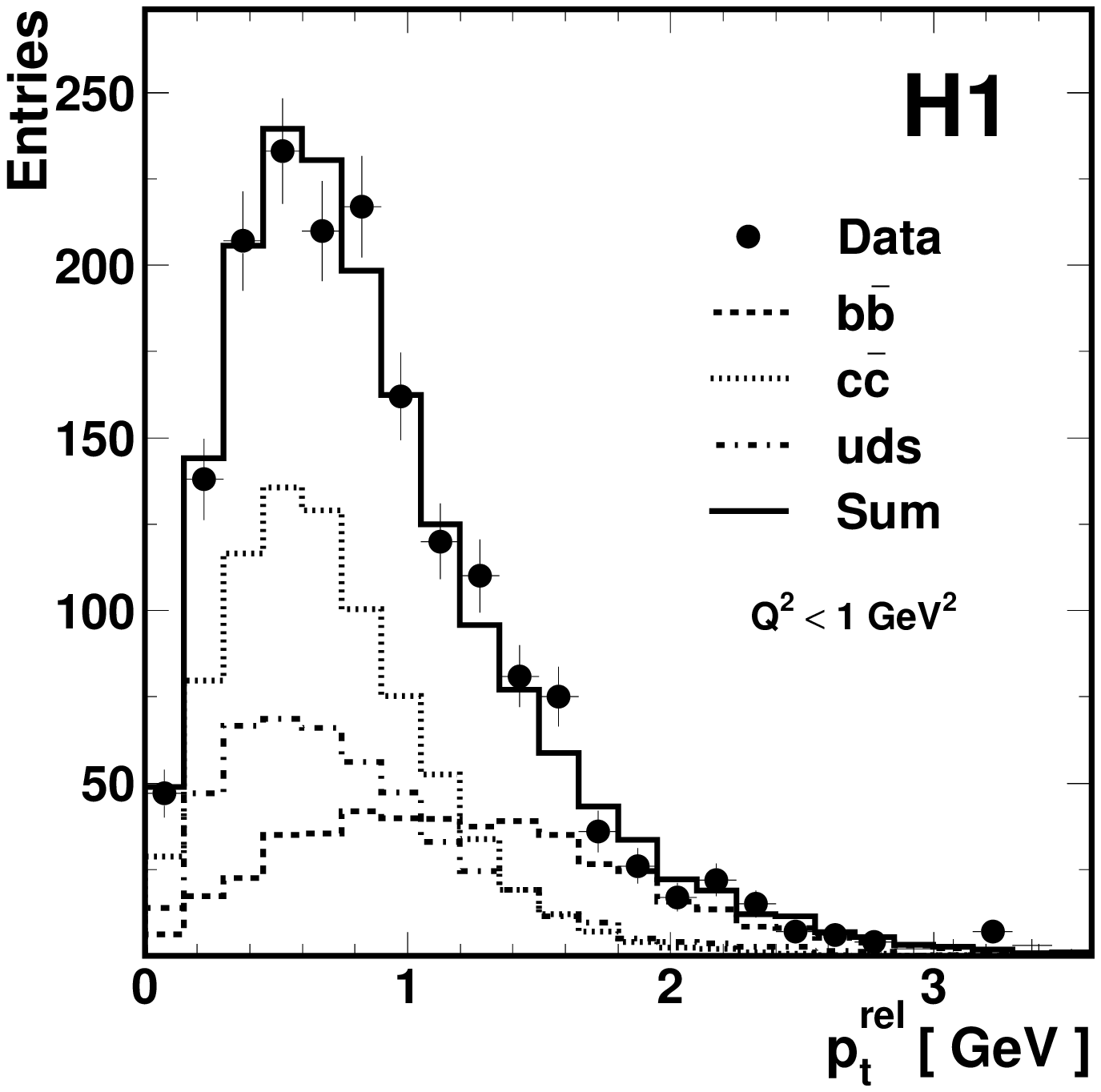,width=9cm}}
\put(2.5,6.7){\large a)}
\put(13.2,6.7){\large b)}
\end{picture}
\caption{Distributions in photoproduction of
a) the impact parameter \del of the muon track and 
b) the transverse muon momentum \ptrel relative
to the axis of the associated jet.
Included in the figure are the
estimated contributions of events arising from 
$b$ quarks (dashed line), \mbox{$c$ quarks} (dotted line)
and the light quarks (dash-dotted line).
The shapes of the distributions of the different sources 
are taken from the PYTHIA Monte Carlo simulation
and their relative fractions are
determined  from a fit to the two-dimensional 
data distribution of \ptrel and \del (see text).}
\label{fig:final-signal}
\end{figure}

\begin{figure}
\setlength{\unitlength}{1cm}
\begin{picture}(14.0,9.)
\put(6.4,7.9){\large \sf Electroproduction}
\put(-.8,-.3){\epsfig{file=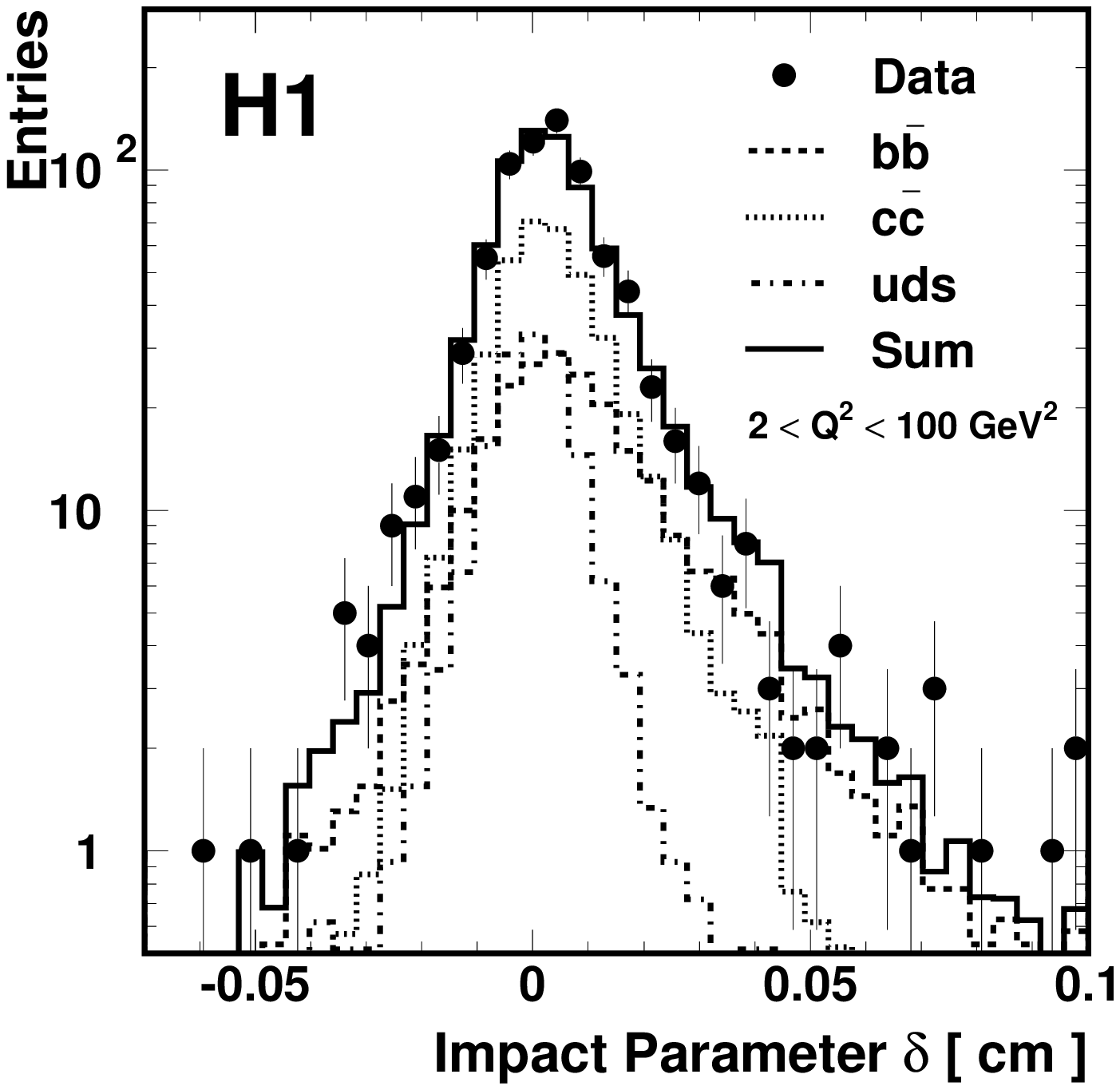,width=9cm}}
\put(7.5,-.3){\epsfig{file=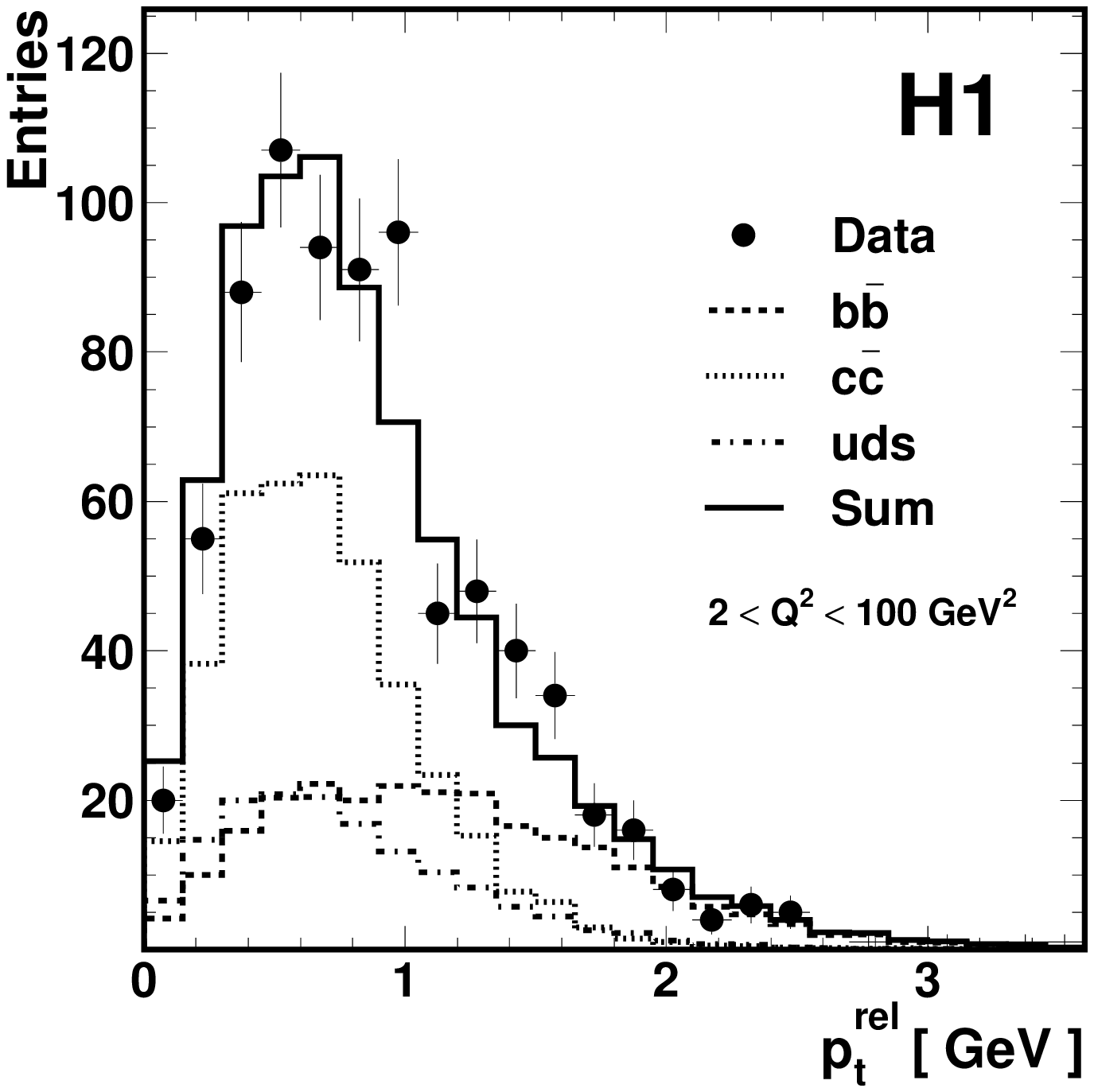,width=9cm}}
\put( 2.5,6.7){\large a)}
\put(13.2,6.7){\large b)}
\end{picture}
\caption{Distributions in electroproduction of
a) the impact parameter \del of the muon track and 
b) the transverse muon momentum \ptrel relative
to the axis of the associated jet.
Included in the figure are the
estimated contributions of events arising from 
$b$ quarks (dashed line), \mbox{$c$ quarks} (dotted line)
and the light quarks (dash-dotted line).
The shapes of the distributions of the different sources 
are taken from the RAPGAP Monte Carlo simulation
and their relative fractions are
determined  from a fit to the two-dimensional 
data distribution of \ptrel and \del (see text).}
\label{fig:dfinal-signal}
\end{figure}

\begin{figure}
\setlength{\unitlength}{1cm}
\begin{picture}(14.0,9)
\put(6.7,7.9){\large \sf Photoproduction}
\put(-.8,-.3){\epsfig{file=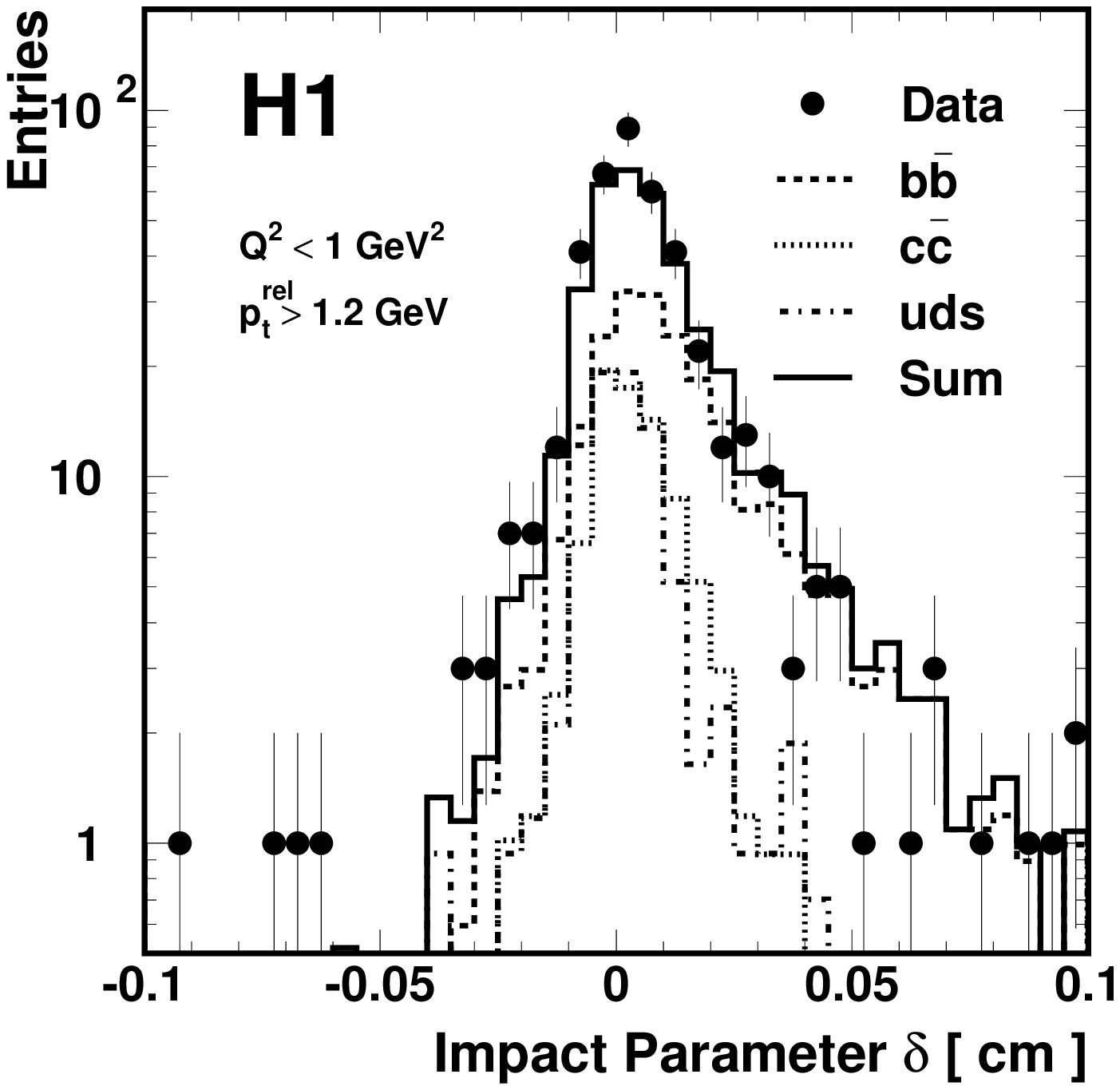,width=9cm}}
\put(7.5,-.3){\epsfig{file=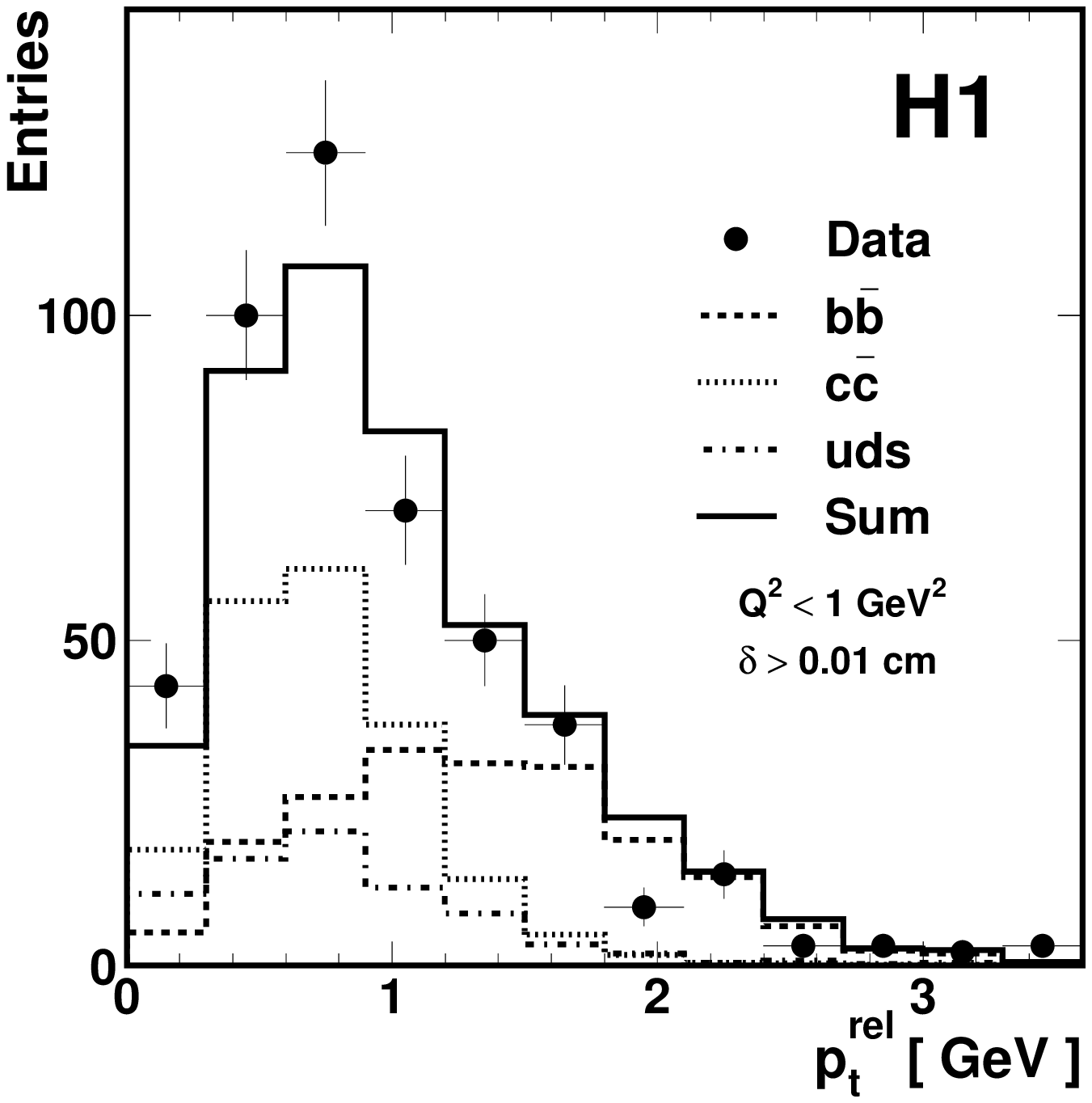,width=9cm}}
\put(2.5,6.7){\large a)}
\put(13.2,6.7){\large b)}
\end{picture}
\caption{Distributions in the restricted photoproduction 
sample of a) the impact parameter \del 
for events with \ptrel$>1.2$ GeV and
b) the transverse muon momentum
\ptrel relative to the jet axis for tracks 
with impact parameter \del $>0.01$ cm. 
The predictions for the contributions to the restricted sample from
$b$ events (dashed line), $c$ events (dotted line) and light 
quark events (dash-dotted line) ,
as determined from a fit to the two-dimensional 
distribution of \ptrel and \del 
in the full data sample (see text), are also shown.
}
\label{fig:final-highp}
\end{figure}

\begin{figure}
\setlength{\unitlength}{1cm}
\begin{picture}(14.0,9)
\put(6.4,7.9){\large \sf Electroproduction}
\put(-.8,-.3){\epsfig{file=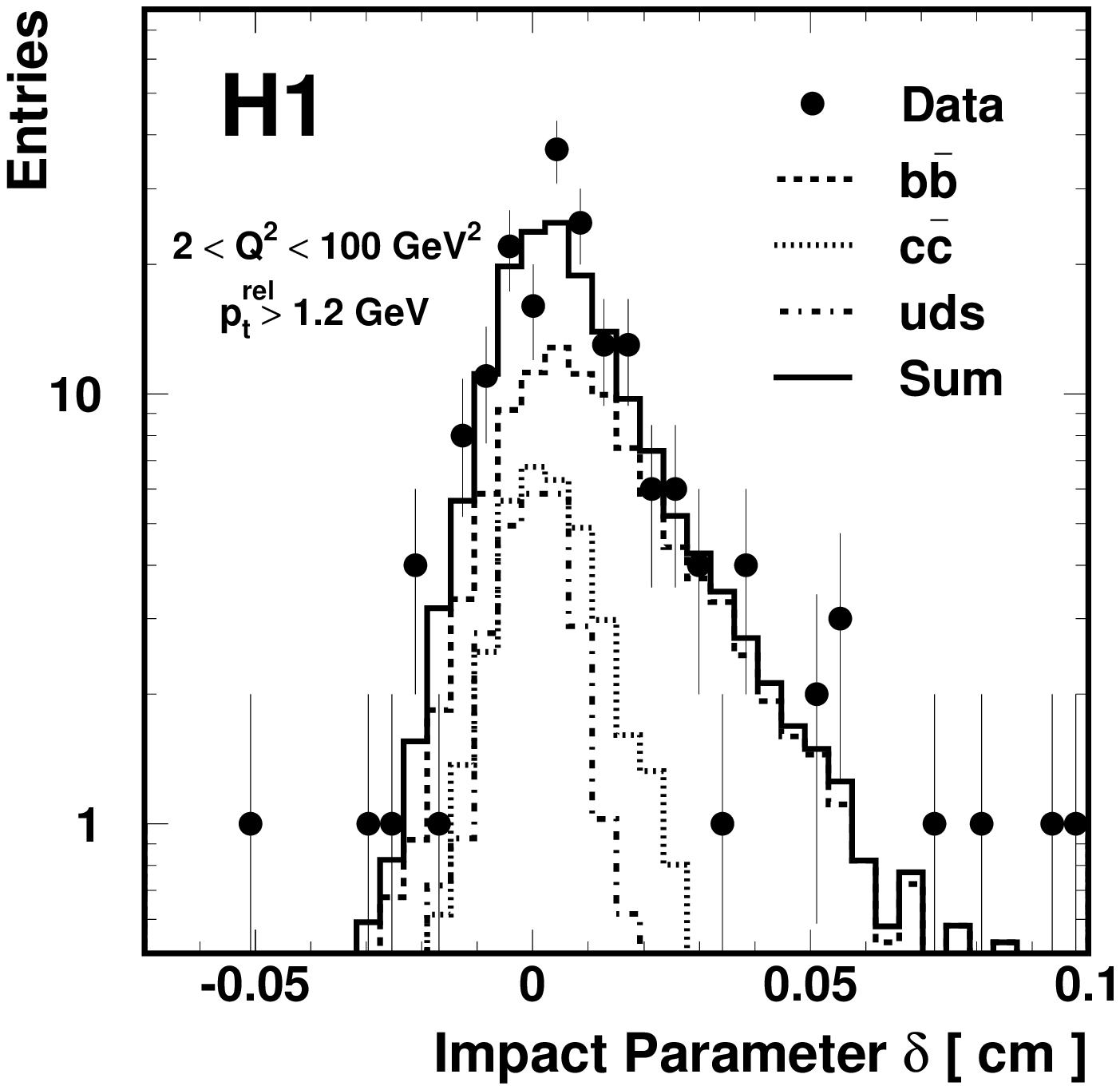,width=9cm}}
\put(7.5,-.3){\epsfig{file=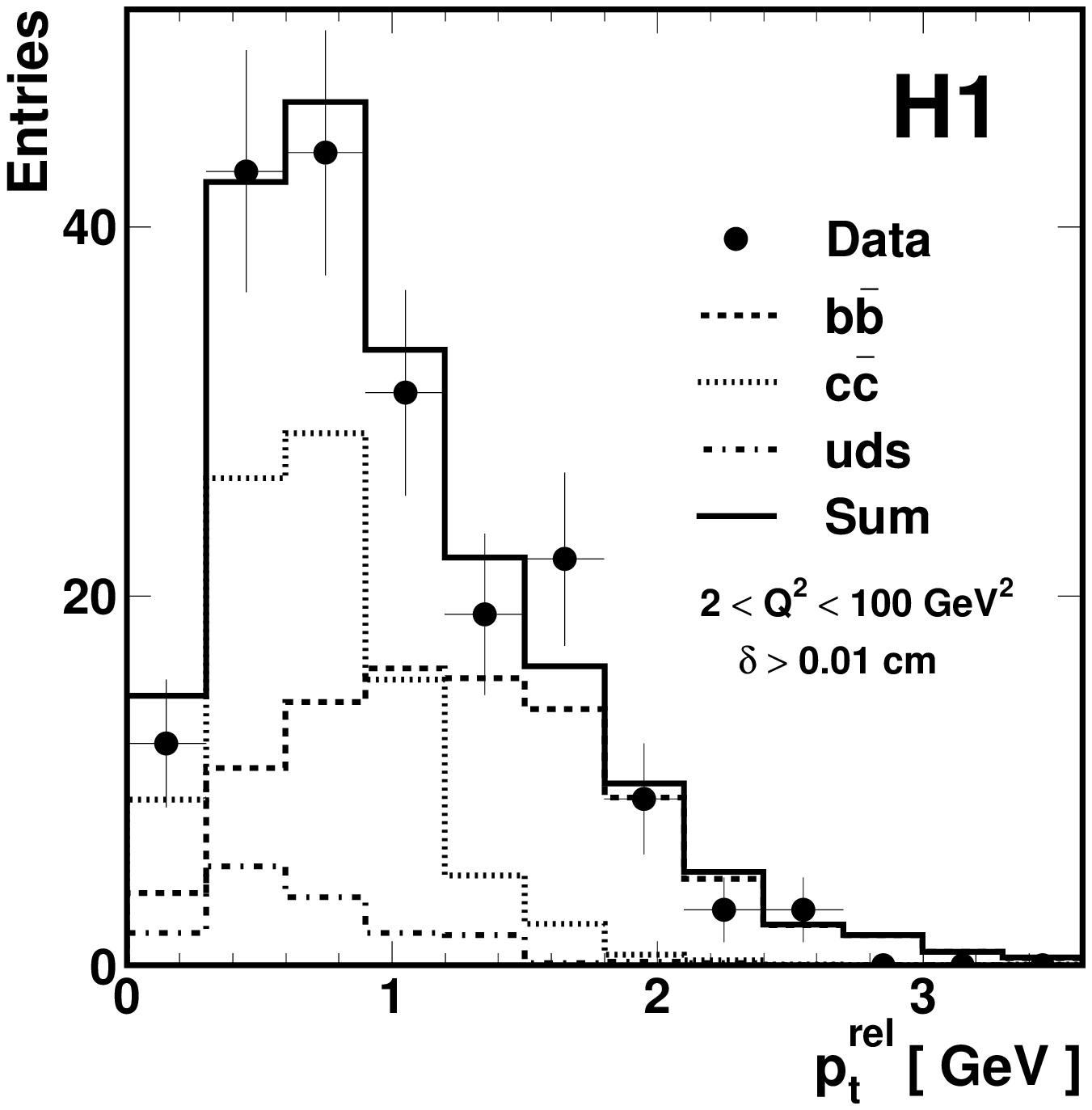,width=9cm}}
\put(2.5,6.7){\large a)}
\put(13.2,6.7){\large b)}
\end{picture}
\caption{Distributions in the restricted electroproduction 
sample of a) the impact parameter \del 
for events with \ptrel$>1.2$ GeV and 
b) the transverse muon momentum 
\ptrel relative to the jet axis for tracks 
with impact parameter \del $>0.01$ cm.
The predictions for the contributions to the restricted sample from
$b$ events (dashed line), $c$ events (dotted line) and light 
quark events (dash-dotted line),
as determined from a fit to the two-dimensional 
distribution of \ptrel and \del 
in the full data sample (see text), are also shown.
}
\label{fig:dfinal-highp}
\end{figure}

\newpage

\begin{figure}
\setlength{\unitlength}{1cm}
\begin{picture}(14.0,16)
\put(6.6,16.3){\large \sf Photoproduction}
\put(-.5,8.){\epsfig{file=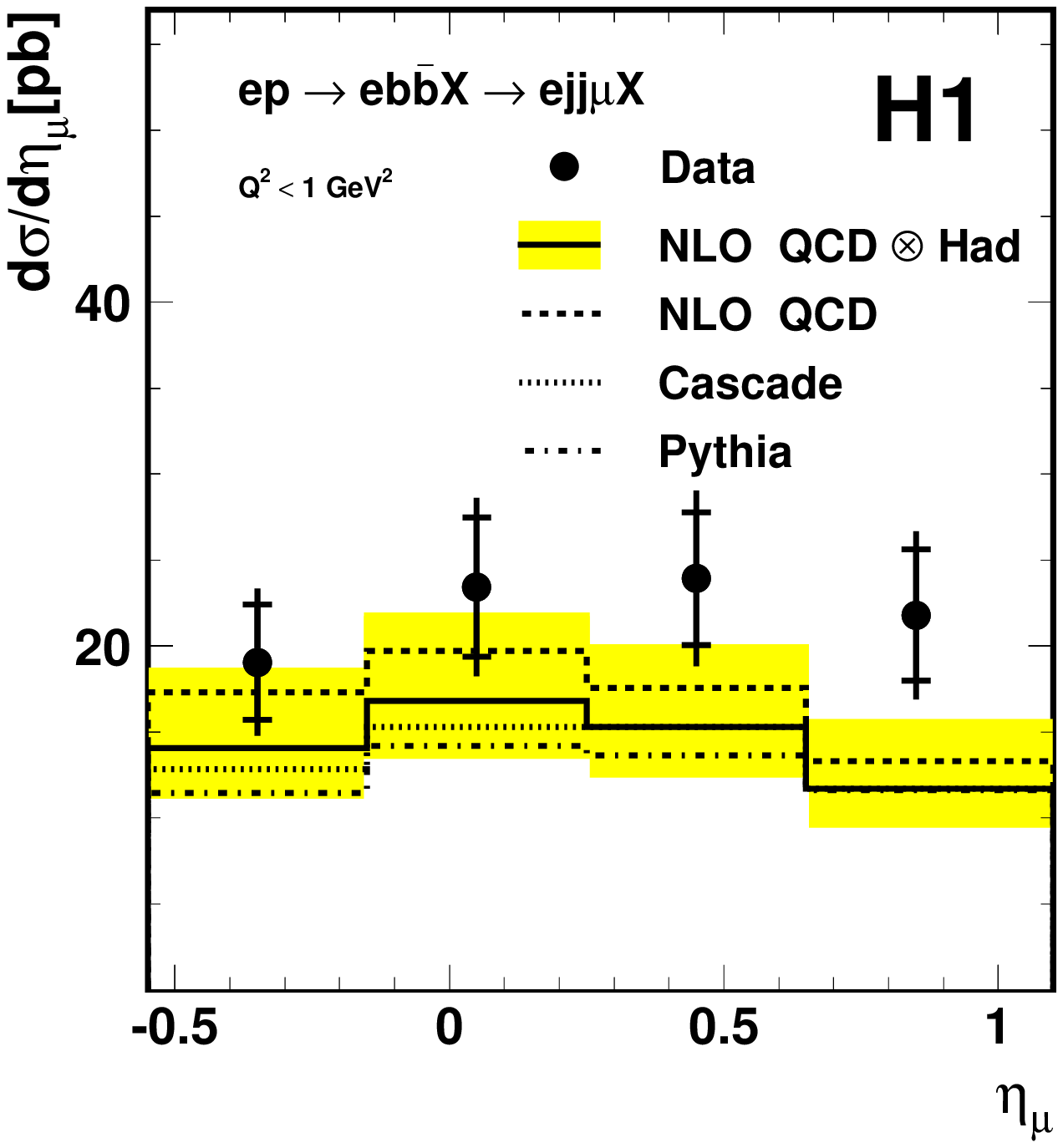,width=9cm}}
\put(7.8,8.){\epsfig{file=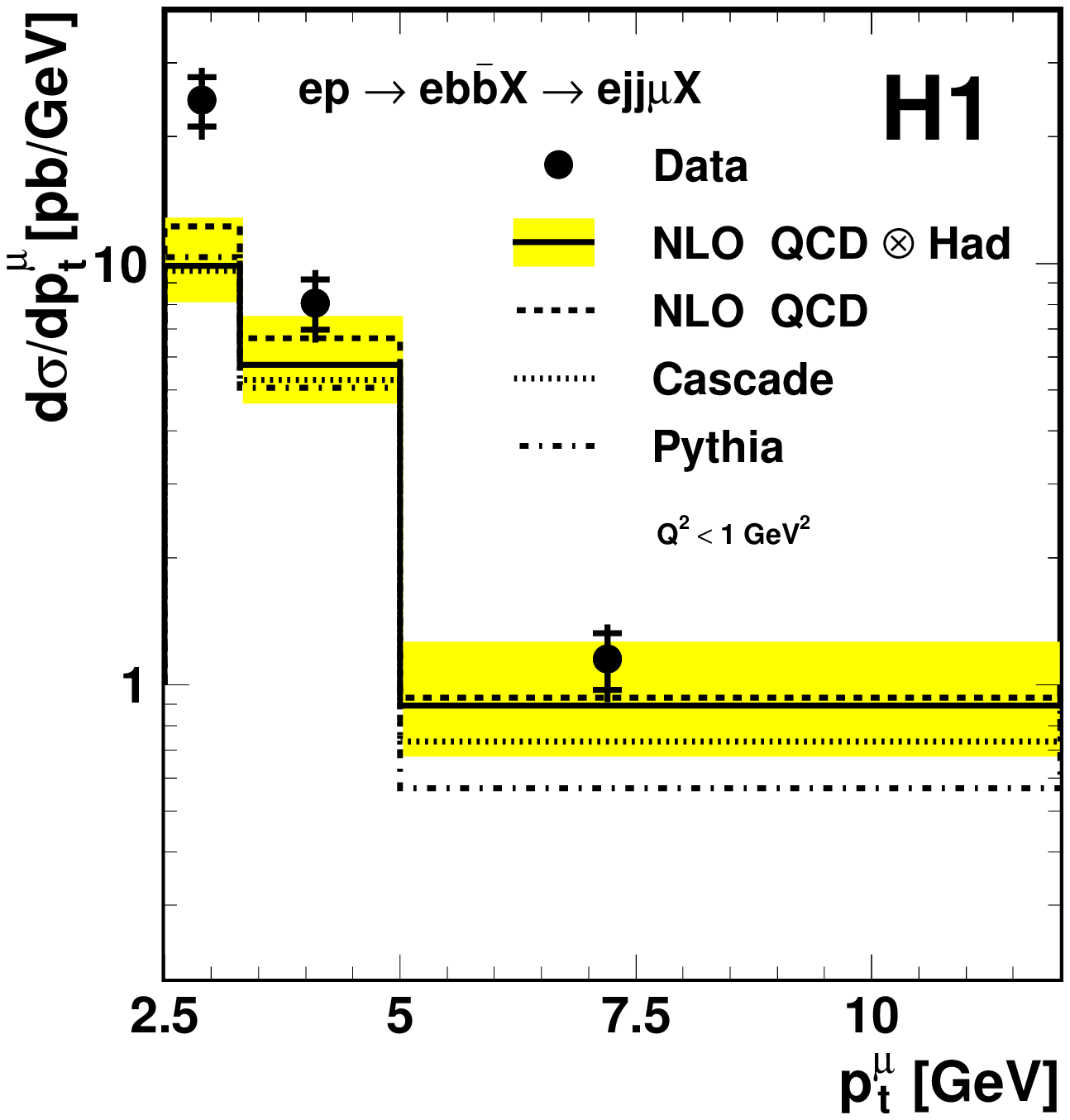,width=9cm}}
\put(-.5,0.){\epsfig{file=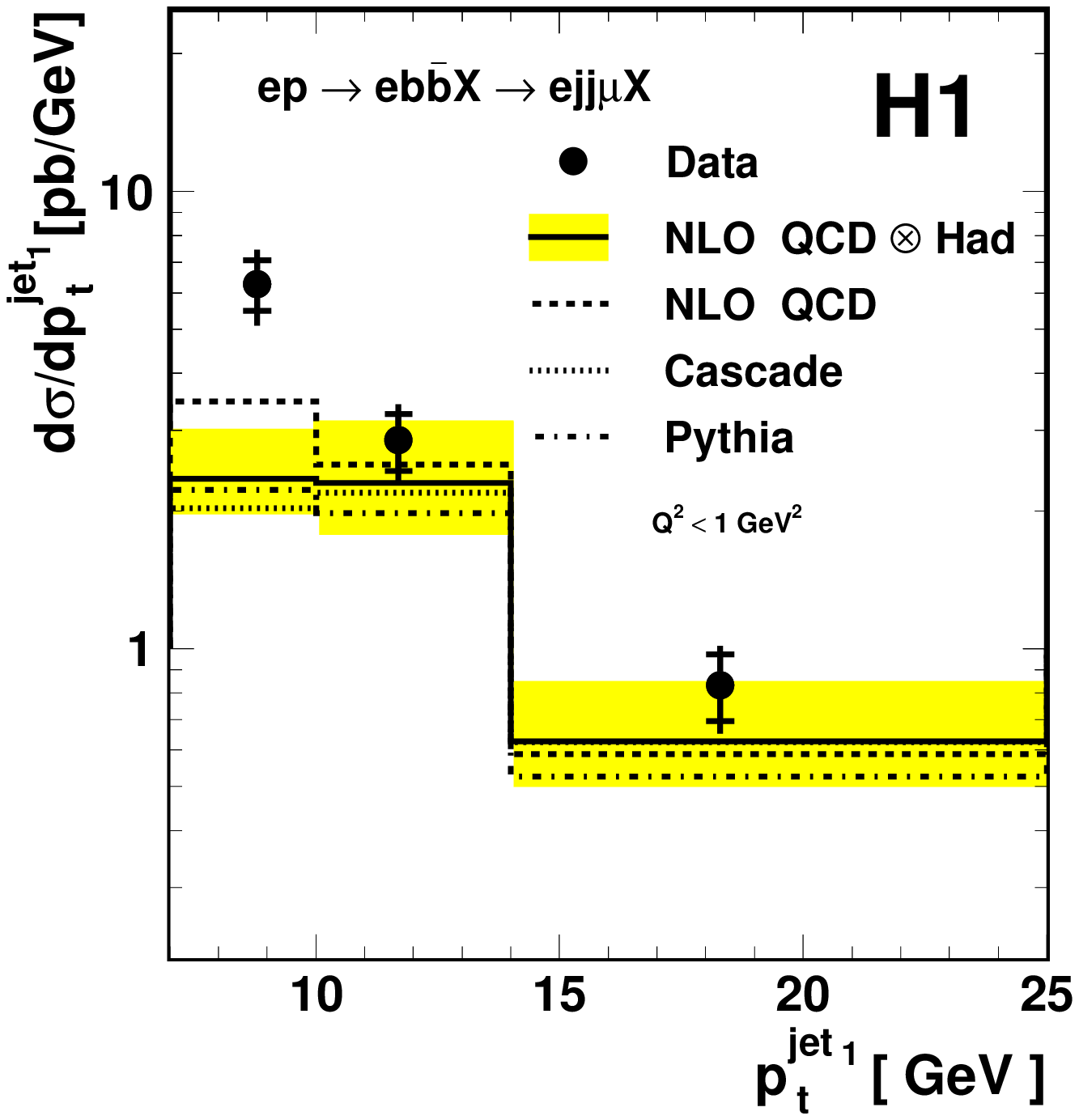,width=9cm}}
\put(7.8,0.){\epsfig{file=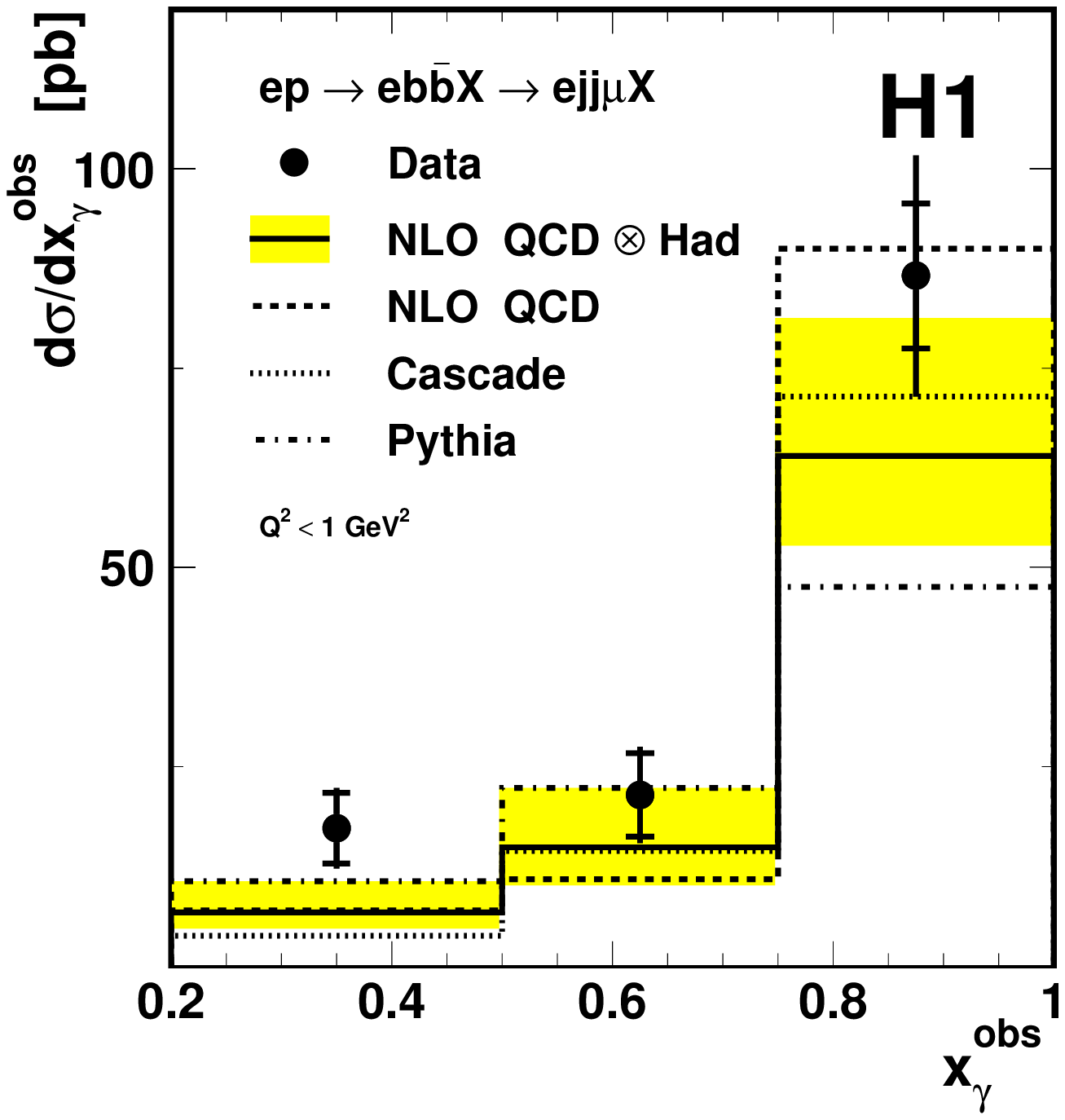,width=9cm}}
\put(1.8,13.8){\large a)}
\put(10.8,14.6){\large b)}
\put(1.8,6.6){\large c)}
\put(13.5,7.0){\large d)}
\end{picture}
\caption{
Differential cross sections for the photoproduction process
$ep \rightarrow e b \bar{b} X \rightarrow e jj\mu X'$ 
in the kinematic range
$Q^2<1$ GeV$^2$, $0.2<y<0.8$, \ptmu $>2.5$ GeV, 
$-0.55<\eta^{\mu}<1.1$,
$p_t^{jet_{1(2)}} > 7(6)$ GeV and $|\eta^{jet}|<2.5$.
The cross sections are shown as functions of
a) the muon pseudo-rapidity $\eta^{\mu}$,
b) the muon transverse momentum $p_t^{\mu}$, 
c) the jet transverse momentum $p_t^{jet_1}$
of the highest transverse momentum jet
and d) the quantity $\xgobsm$.
The inner error bars show the statistical error, 
the outer error bars represent
the statistical and systematic uncertainties added in quadrature.
The NLO QCD predictions at the parton level (dashed line) 
are corrected to the hadron level (solid line)
using the PYTHIA generator. 
The shaded band around the hadron level prediction 
indicates the systematic uncertainties
as estimated from scale variations (see text).
Predictions from the Monte Carlo generator programs 
CASCADE (dotted line) and PYTHIA (dash-dotted line) are also shown.}
\label{fig:xsec-etaptmu}
\end{figure}

\begin{figure}
\setlength{\unitlength}{1cm}
\begin{picture}(14.0,8)
\put(6.7,8.2){\large \sf Electroproduction}
\put(-.5,0.){\epsfig{file=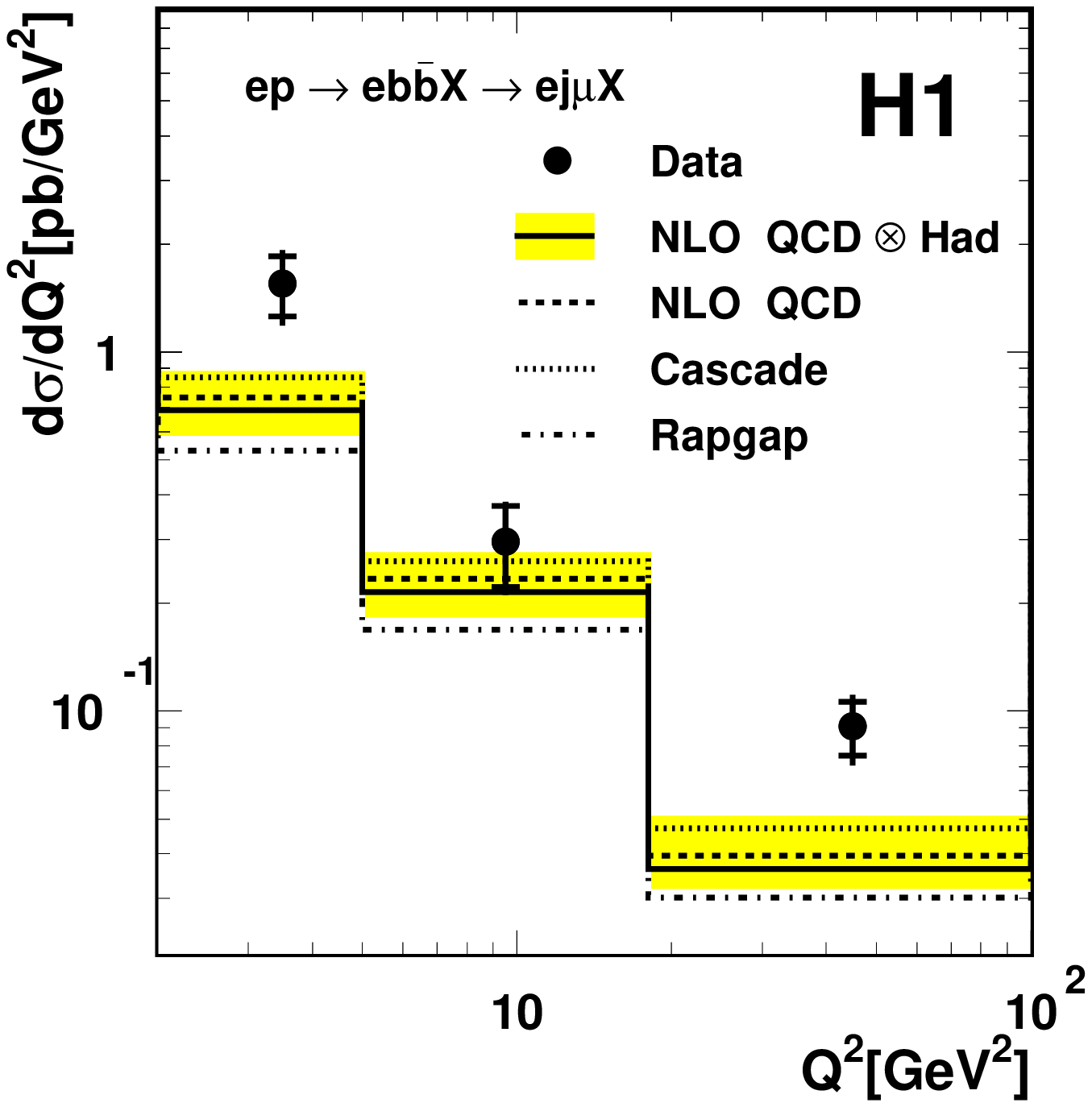,width=9cm}}
\put(7.8,0.){\epsfig{file=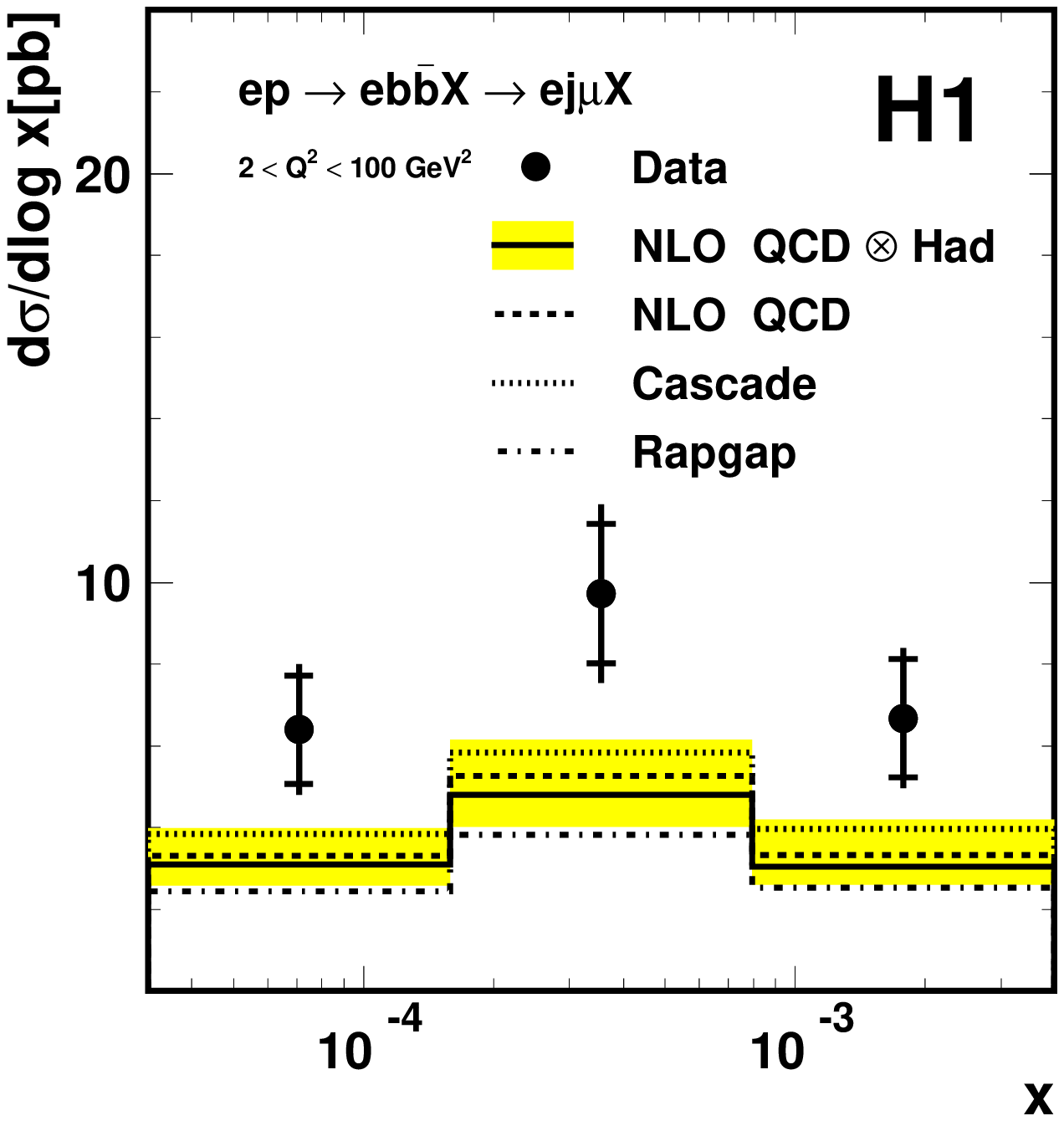,width=9cm}}
\put(1.8,6.7){\large a)}
\put(10.1,5.9){\large b)}
\end{picture}
\vspace{-5mm}
\caption{
Differential cross sections for the electroproduction process
$ep \rightarrow e b \bar{b} X \rightarrow e j\mu X'$ 
in the kinematic range 
$2<Q^2<100$ GeV$^2$, $0.1<y<0.7$, \ptmu $>2.5$ GeV,
\mbox{$-0.75<\eta^{\mu}<1.15$}, 
$p^{Breit}_{t,jet} > 6$ GeV and $|\eta^{jet}|<2.5$.
The cross sections are shown as functions of
a)  the photon virtuality $Q^2$ and b)  
the Bjorken scaling variable $x$.
The inner error bars show the statistical error, 
the outer error bars represent 
the statistical and systematic uncertainties added in quadrature.
NLO QCD predictions at the parton level (dashed line)  are
corrected to the hadron level (solid line)
using the RAPGAP generator. 
The shaded band around the hadron level prediction 
indicates the systematic uncertainty
as estimated from scale variations (see text).
Predictions from the Monte Carlo generator programs 
CASCADE (dotted line) and RAPGAP (dash-dotted line) are also shown.
}
\label{fig:dxsec-q2x}
\end{figure}

\begin{figure}
\setlength{\unitlength}{1cm}
\begin{picture}(14.0,16.)
\put(-0.5,8.){\epsfig{file=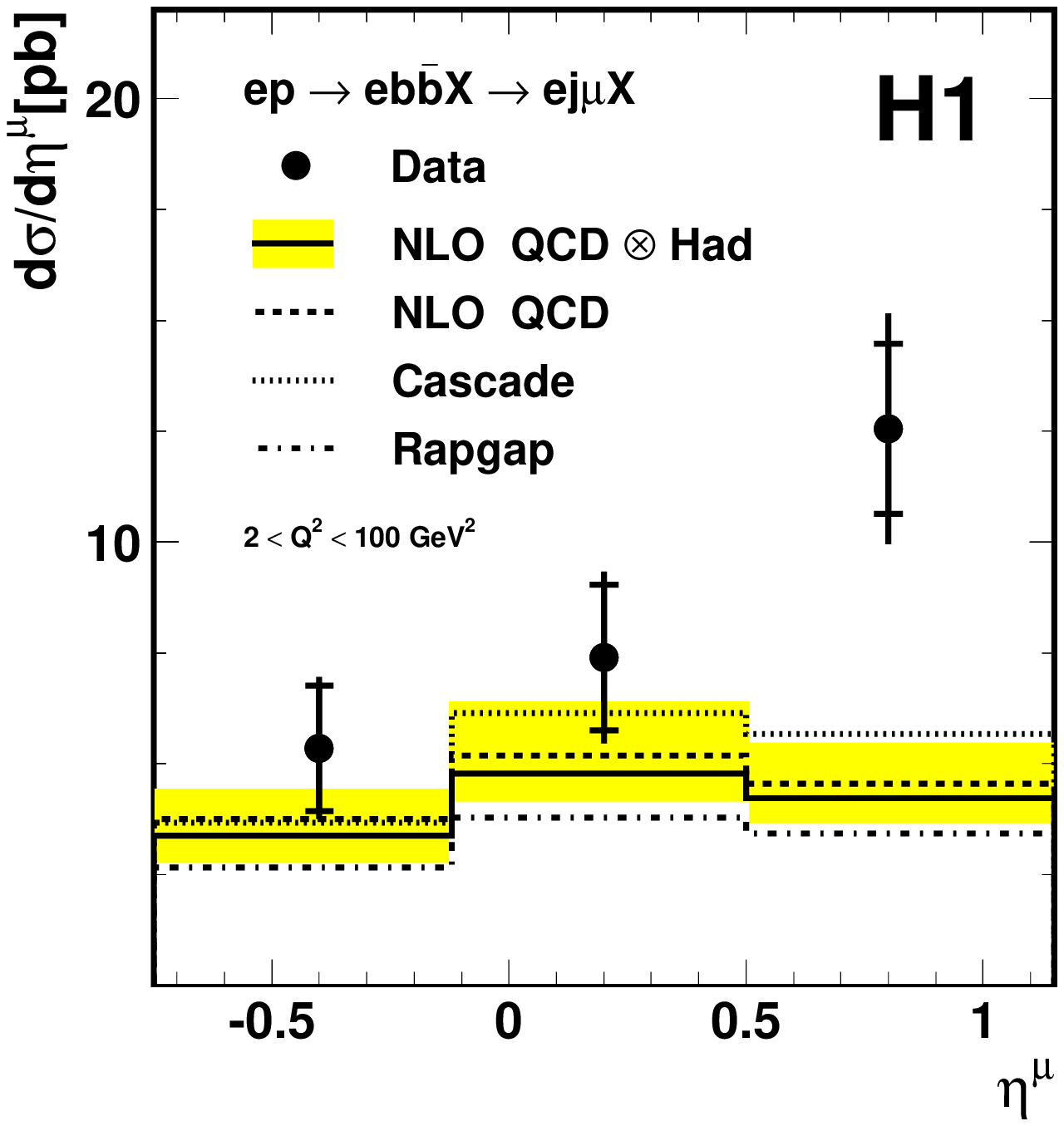,width=9cm}}
\put(7.8,8.){\epsfig{file=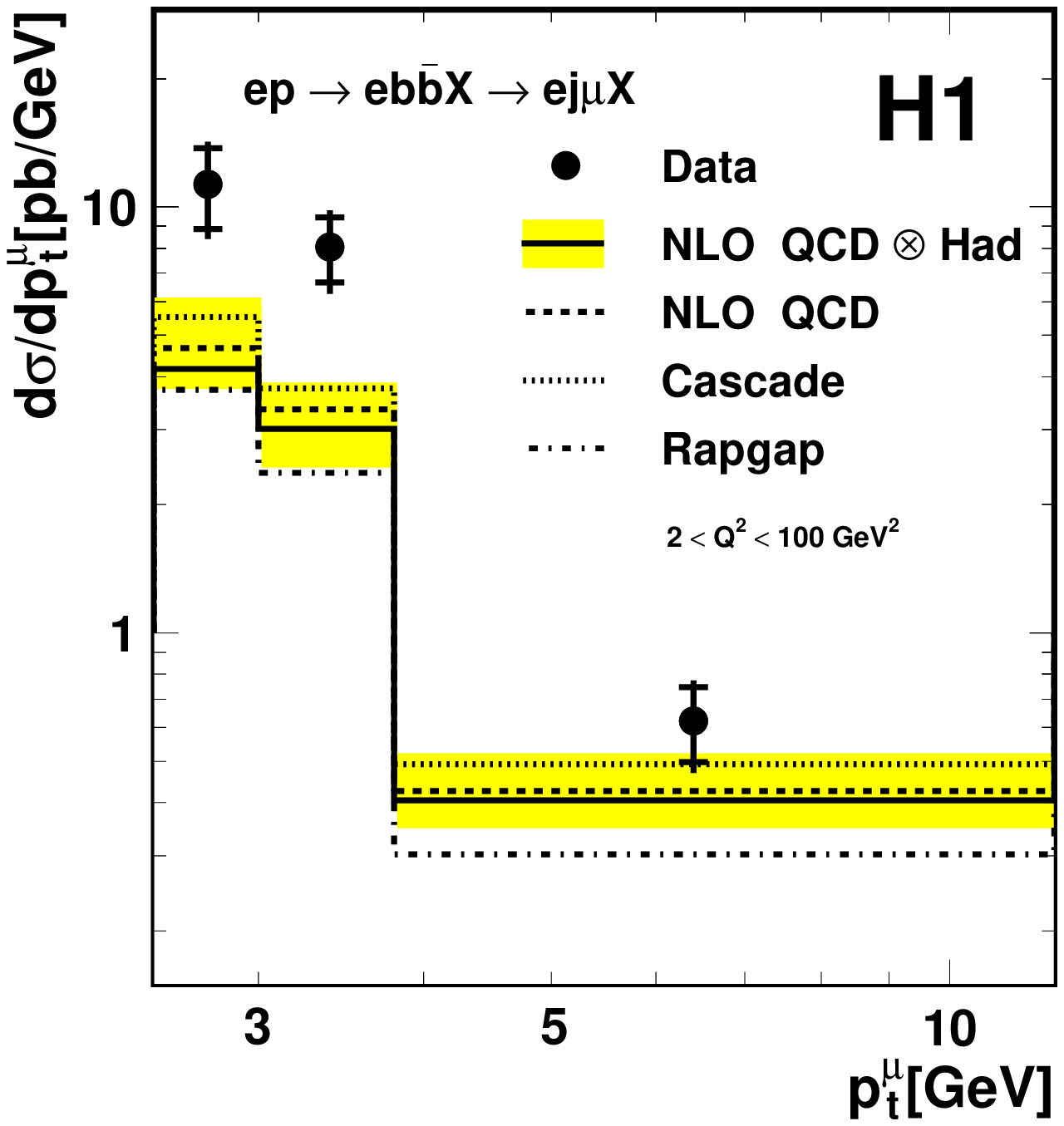,width=9cm}}
\put(-.5,0.){\epsfig{file=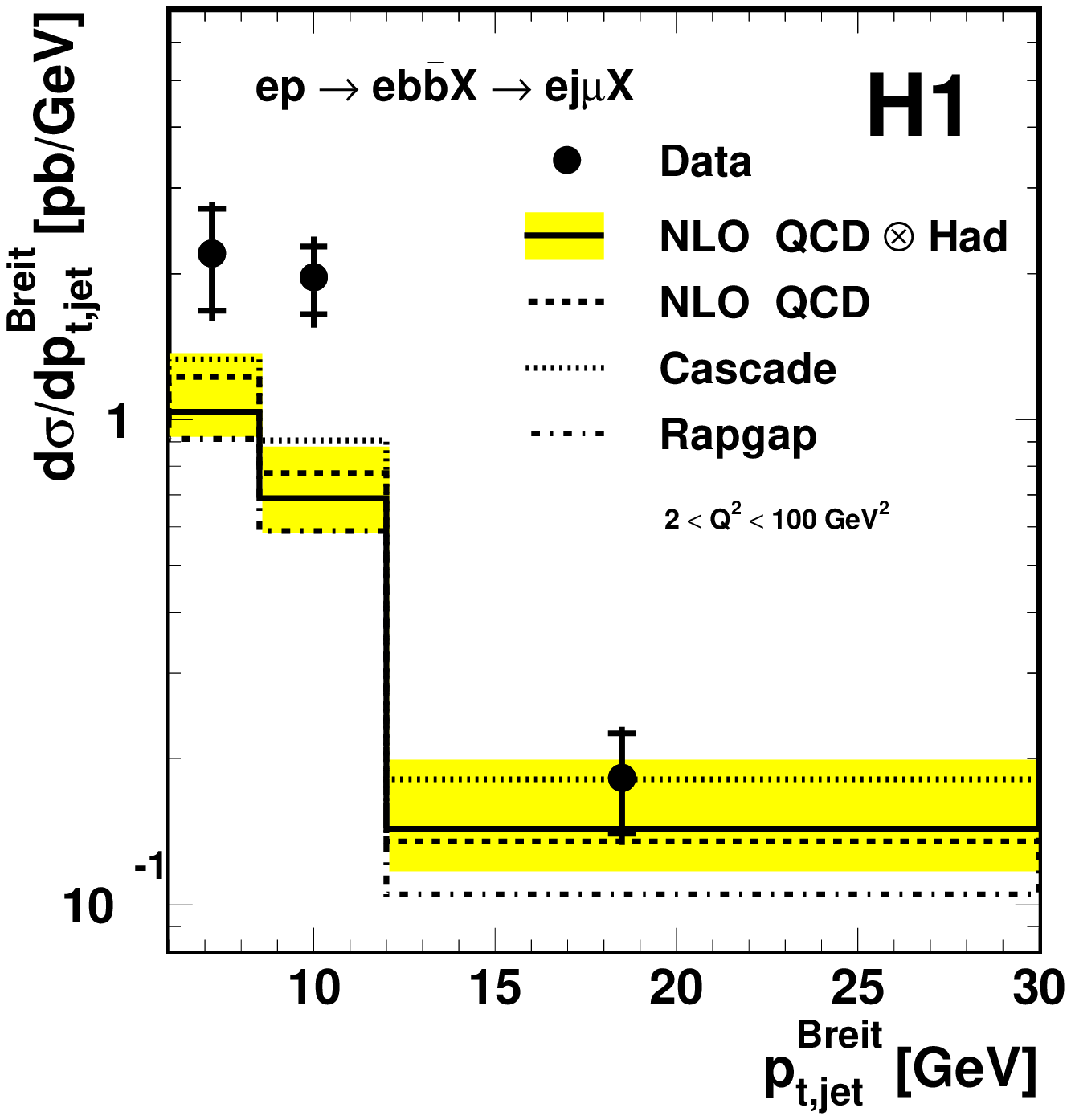,width=9cm}}
\put(6.3,16.3){\large \sf Electroproduction}
\put(6.2,14.4){\large a)}
\put(14.7,13.0){\large b)}
\put(6.4,5.0){\large c)}
\end{picture}
\caption{
Differential cross sections for the electroproduction process
$ep \rightarrow e b \bar{b} X \rightarrow e j\mu X'$ 
in the kinematic range
$2<Q^2<100$ GeV$^2$, $0.1<y<0.7$, \ptmu $>2.5$ GeV,
$-0.75<\eta^{\mu}<1.15$, $p^{Breit}_{t,jet} > 6$ GeV and $|\eta^{jet}|<2.5$.
The cross sections are shown as functions of 
a) the muon pseudo-rapidity $\eta^{\mu}$,
b) the muon transverse momentum $p_t^{\mu}$ and
c) the transverse momentum $p^{Breit}_{t,jet}$ of the
leading jet in the Breit frame.
The inner error bars show the statistical error, 
the outer error bars represent
the statistical and systematic uncertainties added in quadrature.
The NLO QCD predictions at the parton level (dashed line) 
are corrected to the hadron level (solid line)
using the RAPGAP generator. 
The shaded band around the hadron level prediction 
indicates the systematic uncertainty
as estimated from scale variations (see text).
Predictions from the Monte Carlo 
generator programs 
CASCADE (dotted line) and RAPGAP (dash-dotted line) are also shown.
}
\label{fig:dxsec-etaptmu}
\end{figure}

\newpage

\begin{figure}
\setlength{\unitlength}{1cm} 
\begin{picture}(14.0,10.)
\put(-.5,-1.){\epsfig{file=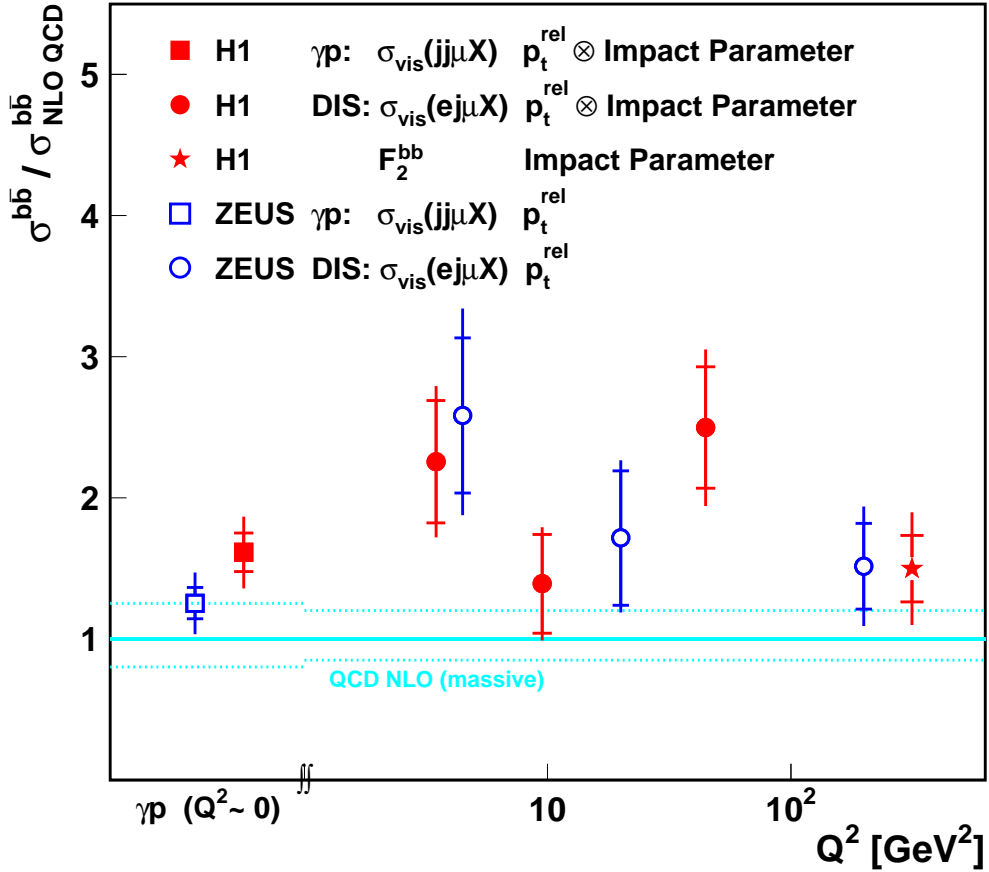,width=16cm}}
\end{picture}
\caption{Ratio of beauty production cross section 
measurements at HERA to NLO QCD predictions.
The results of this paper (solid circles and squares)
are compared with ratios determined using the 
measurements taken from~\cite{Chekanov:2004xy,Chekanov:2004tk,Aktas:2004az}.
The photoproduction points are plotted at different horizontal positions
for better visibility. 
Note that cross section definitions and kinematic ranges are 
somewhat different for the different data points.
The dotted lines indicate the typical theoretical error due to scale 
uncertainties. 
The theoretical prediction used to form the ratio with the measurement
by the ZEUS Experiment~\cite{Chekanov:2004xy}, shown as an
open square, is calculated using the same program and
parameter choices as for the prediction for this measurement (full square).
Different parameter choices, e.g.\,for the modelling of the 
hadronisation and decay of the $B$-hadron, lead to a variation 
of the prediction of $\sim 10\%$.
}
\label{fig:summary}
\end{figure} 
%

\end{document}